\documentclass[journal]{IEEEtran}
\usepackage{graphicx}
\usepackage{float}
\usepackage{amsfonts}
\usepackage{amsmath}
\usepackage{amssymb}
\usepackage{cite}
\usepackage{bm}
\usepackage{mathrsfs}
\usepackage{epsfig}
\usepackage{algorithm}
\usepackage{algorithmic}
\usepackage{amsthm}
\usepackage{color}
\usepackage{graphicx}
\usepackage{caption}
\begin{document}

\title{Application of Compressive Sensing Techniques in  Distributed Sensor Networks: A Survey}
% is not built to handle multiple paragraphs
%

\author{Thakshila~Wimalajeewa,~\IEEEmembership{Senior Member,~IEEE}
        and ~Pramod~K ~Varshney,~\IEEEmembership{Life Fellow,~IEEE}
       \thanks{This material is based upon work supported by the National Science
Foundation under Grant No. ENG 60064237. Thakshila Wimalajeewa is with BAE Systems, Burlington MA. This work was done when she was at Syracuse University, Syracuse, NY.  Pramod K Varshney is with the Department
of Electrical Engineering and Computer Science, Syracuse University, Syracuse,
NY,  USA.  E-mail: thakshila.wimalajeewa@ieee.org, varshney@syr.edu.}% <-this % stops a space
}

%\markboth{Journal of \LaTeX\ Class Files,~Vol.~6, No.~1, January~2007}%
%{Shell \MakeLowercase{\textit{et al.}}: Bare Demo of IEEEtran.cls for Journals}



\maketitle

\begin{abstract}
%Compressive sensing (CS) has been shown to be promising  in  a wide variety of applications including compressive imaging, video processing, communication, and radar to name a few. Out of the many potential applications,
In this survey paper, our goal is to discuss recent  advances  of compressive sensing (CS) based solutions in wireless sensor networks (WSNs) including
the main ongoing/recent research efforts, challenges and research trends in this area.
In WSNs, CS based techniques are well motivated by not only the sparsity prior observed in different forms  but also by  the requirement of  efficient  in-network  processing in terms of transmit power and communication bandwidth  even with nonsparse signals.
In order to apply CS  in a variety of WSN applications efficiently, there are several factors to be considered beyond the standard CS framework.  We start the discussion  with a brief introduction to the theory of CS and then  describe the motivational factors behind the potential  use of CS in WSN applications.  Then, we identify three main areas along which the standard CS framework is extended {so that CS can be efficiently applied to solve   a variety of problems specific to WSNs}. In particular, we emphasize   {on the significance of extending the  CS framework} to (i).  take communication constraints into account while  designing projection matrices  and reconstruction algorithms for signal reconstruction  in centralized as well in decentralized settings, (ii)  solve a variety of inference problems such as detection, classification and parameter estimation, with compressed data without signal reconstruction  and (iii) take practical communication aspects such as measurement quantization, physical layer secrecy constraints, and imperfect channel conditions  into account.  Finally, open research issues and challenges
are discussed in order to provide perspectives for future research
directions.
%With this paper, the readers are expected to gain  a  thorough
%understanding of the potential of CS in solving a variety of  problems in WSNs with high dimensional data under practical communication  constraints.
\end{abstract}

\begin{IEEEkeywords}
 Wireless sensor networks, Data gathering, Distributed inference, Data compression, Compressive sensing (CS), Distributed/decentralized  CS,  Fading channels, Physical layer secrecy, Compressive detection, Compressive classification, Quantized CS
\end{IEEEkeywords}

\section{Introduction}
Over the last two decades, the wireless sensor network (WSN) technology has  gained increasing attention by both the research community and actual users \cite{Akyildiz_CN2002,Mainetti_SoftCOM2011,
Stankovic_CSN2011,Rawat_JoS2014,Rashid_JNCA2016}. Applications of WSNs  span  a wide range including   environmental monitoring and surveillance \cite{Othman_EP2012,Raghunathan_MCOM05}, detection and classification \cite{Chamberland_SPM2007,Jiang_TWC05,Wimalajeewa_TWC2008,Duarte_JoPDC2004}, target/object tracking \cite{Zuo_icassp2007,Ozdemir_tsp2009,saber_ACC2007,zou_TMC07,djuric_tsp2008}, industrial applications \cite{Flemmini_MJ2009,Gungor_IE2009}, and health care \cite{Alemdar_CN2010} to name a few.   {In addition to domain specific and task-oriented applications, WSN technology has  been identified as one of the key components in designing future Internet of Things (IoT) platforms \cite{Khan_COMT2016,Oteafy_MCOM18}.}

A typical sensor network  consists of multiple  sensors of the same or different modalities/types  deployed over  a geographical area for monitoring a phenomenon of interest (PoI). Once deployed, the distributed sensors are required to form a connected network without a backbone infrastructure as in cellular networks.  Most of the sensors are power constrained since they are equipped with small sized batteries which are difficult or impossible to be replaced especially in hostile environments. At the same time, the available (limited) communication bandwidth needs to be efficiently used while  exchanging  information for efficient  fusion. Thus, sensor networks are inherently resource constrained and they starve for energy and communication efficient  protocols \cite{Akyildiz_CN2002,Pantazis_COMT2013}. While distributed sensor fusion under resource constraints  has been a research topic investigated for decades,   the emergence of new sensors of different modalities  that are capable of  generating
huge amounts of data in heterogeneous environments  makes   real-time fusion increasingly challenging \cite{Atat_IET2017,Wu_STST2016,Wu_STST2016_2}. Thus, desirable (or lossless) data compression is very important in  designing WSNs for task-oriented as well as IoT based information systems.

%
% There is abundant literature related  to energy-savings in
%WSNs that has appeared  in the
%last several  years.
%
%There are also several  survey and tutorial papers that discuss energy efficiency in WSNs, e.g.,   \cite{Pantazis_COMT2013}.

%
%One of the main  objectives of WSN research is to design  efficient and scalable protocols and algorithms  to perform a given task (or multiple tasks)  for such applications taking  the inherent resource constraints into account.   However, there is still much ongoing research on investigating how to optimize power and communication bandwidth  in resource constrained wireless sensor
%networks for task-oriented as well 'smart' sensing applications such as IoT \cite{}.

Advances in compressive sensing (CS) have  led to novel ways of thinking about approaches to design energy efficient WSNs {with low cost data acquisition}. CS has emerged as a promising  paradigm for efficient high dimensional sparse signal  acquisition.
%Sparsity is one of the low dimensional structures exhibited in many signals of interest including audio, video and radar signals \cite{Eldar_B1}.
In the CS framework,  a high dimensional  signal  can be reliably recovered with a small number of random projections under certain conditions if the signal of interest is sufficiently sparse  \cite{candes_TIT06,candes_SPM08,donoho_TIT06,candesTao_TIT06}. In particular, compression is a simple linear operation implemented using random projection matrices which is independent of the signal parameters. In order to reconstruct the original high dimensional signal from its compressed version, several reconstruction techniques have been proposed where each one is different from the other in terms of their  recovery performance and  computational complexity \cite{Beck_SIAM2009,tropp_TIT2007,Blumensath_ACHN2009,Eldar_B1}.
%Since its inception over a decade ago,  CS theory has been successfully applied to a wide variety of applications including compressive  imaging, video processing,  wireless communication, cognitive radio networks, sensor networks,   radar signal processing  to name a few.

  CS is well motivated for a variety of WSN applications due to several reasons.   Due to inherently  scarce  energy and communication  resources in WSNs, data compression prior  to transmission within WSNs is vital. On the other hand, sparsity is a common characteristic of many signals of interest  that can be observed  in various  forms/dimensions.
 % Thus, CS is readily applicable for sparse signal acquisition  with only a small number of measurements compared to the signal dimension.
%In such applications, the data observed at individual nodes is reproduced in a centralized or distributed manner  by letting sensors transmit only a small amount of (compressed) information.
  Thus,  an immediate use  of CS in WSNs is  {data gathering with reduced rate samples}, as required by many environment  and infrastructure  monitoring applications. CS based data gathering   may either exploit  temporal, and/or  spatial  sparsity. Signal reconstruction with compressed data in CS was  initially developed for a single measurement vector (SMV) which  was later extended to estimate multiple sparse signals sharing joint  structures using  multiple measurement vectors (MMVs) \cite{Tropp3,Tropp4}.  Direct use   of CS with SMV or MMVs may not be desirable  due to  communication constraints and specific application requirements in large scale WSNs.  {In particular, recent extensions and modifications  of CS    to cope with  communication/energy constraints and variations of sensor readings can be exploited to better utilize CS based techniques in WSNs. These extensions/modifications beyond the standard SMV/MMV  include   the design of  adaptive and sparse projection matrices to compress data at distributed nodes while meeting the desired communication constraints, and distributed and decentralized solutions for signal reconstruction considering different network models.}
%  Thus, extensions of  CS based signal reconstruction techniques  to the distributed and decentralized setting are needed.

 The simple and universal low rate data acquisition scheme  provided by CS enables the design of new approaches  to solve   a variety of inference problems by suitable  fusion of WSN data. In solving inference problems with compressed data,  complete signal reconstruction, as employed  in the standard CS framework, may not be necessary. Instead, constructing a decision statistic directly in the compressed domain is sufficient to make a reliable inference decision, for example,  in intruder detection, early detection of natural disasters in smart environments,  estimation of parameters such as energy radiated by cell stations in  smart cities, and object tracking. Moreover, when applying CS  techniques to perform  different tasks  in WSNs, their  robustness in the presence of issues     such as fading channels, physical layer secrecy concerns and quantization needs to be understood. {This is because, the desirable conditions need to be satisfied  by the standard CS framework can be violated under such practical aspects.  }  Thus,  to make CS ideas practically implementable for  different tasks in a variety of WSN applications, above mentioned  factors beyond the standard CS framework need to be understood. Over the last several years, there has  been extensive  research efforts in this direction.

\subsection{Overview of the Current Paper}
{The goal of this review paper is to discuss in some detail how  the extensions and modifications done to the original CS framework   can be utilized to solve a variety of problems in WSNs under practical considerations. }
{Our  review is based on  the following classification of existing work. We believe that this classification allows us to gather most of the recent modifications/extensions to the CS framework to meet WSN specific objectives  and  would provide the reader a comprehensive understanding on the use  of CS in WSN specific applications. In particular, our discussion covers:  }
\begin{enumerate}
\item [i)]  { Extensions  of the CS framework to operate under communication constraints for data gathering considering}
\begin{itemize}
\item  { form of  sparsity exploited:  temporal, spatial, spatio-temporal}
\item  { data acquisition/collection   techniques: sparse, adaptive, and structured projection matrices, single-hop and multi-hop data collection}
    \item  {reconstruction techniques: centralized, and decentralized implementation of different reconstruction algorithms including optimization based, greedy/iterative and Bayesian algorithms}
    \end{itemize}
\item[ii)]   {Extensions of the  CS framework to solve  a variety of  inference problems without signal reconstruction including}
    \begin{itemize}
    \item  {detection, classification,  parameter  estimation, source localization, and sensor management}
    \end{itemize}
    \item[iii)]     {    Incorporation of  practical communication issues  into the CS framework including}
    \begin{itemize}
     \item  {channel fading}
     \item  {physical layer secrecy constraints and }
       \item  {quantization}
            \end{itemize}
      \item[iv)] { Lessons learned and future directions.}
\end{enumerate}

{In the following, we discuss the most related existing review/survey papers and highlight the contribution of the current paper compared to  the existing papers. }

\subsection{Comparison with Related Survey/Review  Articles}
 CS ideas have  gained  significant interest in a variety of applications such as   imaging \cite{Duarte_SPM2008,Lustig_SPM2008}, video processing   \cite{Lustig_SPM2008},  \cite{Jiang_Bell14,Pudlewski_COMT2013}, cognitive radio networks \cite{Sharma_COMT16,Qin_COM2017}, machine-type communications \cite{Yu_wcnc2017},  radar signal processing \cite{Liang_Globecom10}, and  physical layer operations in  communication systems such as channel estimation in wireless networks \cite{Berger_SPM2010,Peng_2010,Mohammadian_COML2017,Cheng_COM2013,Eltayeb_COM2014,Choi_COM2015},  channel estimation in power line communication \cite{Ding_JSAC2016}, to name a few.  In early review papers/book chapters related to CS, theory, algorithms  and general  applications of CS have been discussed \cite{candes_SPM08,Eldar_B1,Gilbert_Proc2010}. There are also few recent survey papers that discuss recent advances in CS algorithms \cite{Abo-Zahhad_J15,Zhang_Access15}.

%We would like to emphasize that,  there are also several survey/tutorial papers on CS based techniques  in WSNs  covering different aspects.

\begin{table*}[!t]
\renewcommand{\arraystretch}{1.3}
 %if using array.sty, it might be a good idea to tweak the value of
% \extrarowheight as needed to properly center the text within the cells
\caption{Related survey/review papers}
\label{table_relatedpapers}
\centering
% Some packages, such as MDW tools, offer better commands for making tables
% than the plain LaTeX2e tabular which is used here.
\begin{tabular}{|l|l|l|l|}
\hline
Aspect & References & Year &  Contributions  \\
\hline

%\hline
CS Recovery  algorithms  & \cite{Zhang_Access15} & 2015& A survey on sparse signal recovery algorithms with a single measurement vector; \\
with SMV/MMV & $~$ & $~$ & discusses different types of reconstruction algorithms along with a comparative study\\
$~$ & \cite{Rakotomamonjy_SP2011}& 2011&  A survey on sparse signal recovery algorithms with multiple  measurement vectors;\\
$~$ & $~$ & $~$ &discusses optimization and greedy/iterative  based simultaneous sparse approximation \\
$~$ & $~$ & $~$ & algorithms considering the JSM-2 model \cite{Baron_2006} \\
$~$ & \cite{Sundman_JSAN2014} & 2014 &  A review on  CS reconstruction algorithms;  discusses CS reconstruction algorithms \\
$~$ & $~$ & $~$ &  for different distributed network models \\
\hline
CS for  communications and networks & \cite{Qaisar_13} & 2013 & A survey on theory and applications of CS;   discusses compressed data gathering,  \\
$~$ & $~$ & $~$ & distributed compression  and source localization under CS in communications and networks\\
\hline
%Image/video compression & \cite{Pudlewski_COMT2013} & A tutorial  on a distributed compressive video sensing; discusses advantages of  \\
%$~$ & $~$ & CS based video processing vs traditional techniques in a distributed manner\\
%\hline
CS for wireless communication & \cite{Choi_COMT2017} & 2017 & A survey on factors to be considered when applying CS for  channel estimation, interference  \\
$~$ & $~$ & $~$ & cancellation, symbol detection, support identification in wireless communication\\
\hline
Compression techniques in WSNs & \cite{RAZZAQUE_ACM13} & 2013 &A survey on compression techniques used in WSNs for data gathering; compares  \\
$~$ & $~$ & $~$ &   the use of CS based techniques and the conventional compression schemes such as \\$~$ & $~$ & $~$ &  transformed based and distributed source coding \\
\hline
CS for WSNs & \cite{Balouchestani_MWN2011} & 2011&  A survey on CS for WSNs;  discusses the improvements in factors such as
lifetime, delay, \\
$~$ & $~$ & $~$ &  cost and power\\
%\hline
%Recovery algorithms for WSNs & \cite{Sundman_JSAN2014} & A review on  CS reconstruction algorithms;  discusses CS reconstruction algorithms \\
%$~$ & $~$ &  for different distributed network models \\
\hline
CS for image/video data  & \cite{Pudlewski_COMT2013} & 2013&  A tutorial  on a distributed compressive video sensing; discusses the  advantages of  \\
compression in WSNs & $~$ & $~$ & CS based video processing vs traditional techniques in a distributed manner\\
\hline
CS for cognitive radio networks   & \cite{Sun_TWC13} & 2013&  A survey on wideband spectrum sensing techniques; discusses Nyquist and  \\ & $~$ & $~$ &  sub-Nyquist (CS based) techniques for spectrum sensing\\
$~$   & \cite{Sharma_COMT16} & 2016&  A survey on application of CS in cognitive radio networks; discusses CS based wideband  \\ & $~$ & $~$ &  spectrum sensing,   CS based CR-related parameter estimation and \\ & $~$ & $~$ &  the use of CS for radio environment map construction\\
$~$   & \cite{Salahdine_J16} & 2016&  A survey on CS based techniques for cognitive radio networks; discusses the use of CS   \\ & $~$ & $~$ &  for a variety of CR applications including spectrum sensing, channel estimation,\\
& $~$ & $~$ & and multiple-input multiple-output based CR\\
%Data compression for WSNs & \cite{Razzaque_Sen2014} & A survey on different data compression schemes for WSNs; compares CS based approach \\
%$~$ & $~$ &   with other traditional techniques such as \\
\hline
\end{tabular}
\end{table*}

 Applications of  CS in WSNs have been discussed to some  extent in several related papers. There are  survey/review papers available in the literature on CS in communication systems in general where sensor networks are  treated as one application and  some results can be easily applied   to sensor networks as special cases.   In \cite{Qaisar_13}, the application  of CS for compressed data gathering, distributed compression  and source localization has  been  briefly   reviewed  under the general topic of CS for  communications and networks. {Similarly,  the use of CS for communication networks has been reviewed in \cite{Huang_axv13} focusing on getting different physical, network and application layer tasks done. Specific to WSNs, several  topics such as  compressed data gathering exploiting temporal and spatial sparsity, and compressed data  routing in a centralized setting are reviewed in \cite{Huang_axv13}}. In a recent survey paper \cite{Choi_COMT2017}, the authors have emphasized  on the factors to be considered when applying CS for  channel estimation, interference  cancellation, symbol detection, support identification in wireless communication as applicable to different application scenarios. Some of the operations discussed in \cite{Choi_COMT2017} such as simultaneous sparse signal recovery using MMVs  and source localization   are applicable to WSNs as well.

When considering existing survey papers that specifically focus on WSNs, most of them discuss how CS can be utilized for  compressed data gathering  using the centralized architecture. {In a centralized setting, the direct use (by direct use, we mean that there is  no additional work  done on designing projection matrices and/or  reconstruction algorithms  beyond the standard CS framework)  of CS  reduces to the MMV problem if temporal sparsity is exploited and the SMV problem if spatial sparsity is exploited}.    In \cite{RAZZAQUE_ACM13}, {the direct use  of CS for data gathering exploiting spatial sparsity has been considered. A comparative study on the advantages and disadvantages of CS based compression with inter- and intra-signal correlation compared to conventional  compression schemes  employed for WSNs has been  presented. In   \cite{Balouchestani_MWN2011},  the direct use of CS exploiting   spatial sparsity of data collected at multiple nodes for distributed compression has been  considered. The authors have discussed  as to how  certain sensor network parameters such as
lifetime, delay, cost and power can be improved using CS.
 However, the work related to the exploitation of other forms of sparsity, incorporation of communication related issues when  designing projection matrices  and reconstruction techniques for data gathering is missing in \cite{RAZZAQUE_ACM13,Balouchestani_MWN2011}}.  In  \cite{Sundman_JSAN2014, Rakotomamonjy_SP2011}, work related to development of reconstruction algorithms taking different network models and communication architectures has been reviewed. In particular, the authors in  \cite{Sundman_JSAN2014} have discussed the  distributed  development of some reconstruction algorithms for several signal models,  and  network topologies.  Simultaneous sparse approximation algorithms reviewed  in \cite{Rakotomamonjy_SP2011} are applicable for WSNs with the JSM-2 model (as defined in \cite{Baron_2006} and discussed in detail in Section \ref{sub_Sec_MMV_cent}). {While some of the algorithms discussed in this review paper for data gathering have  an overlap to some  extent with the ones that are discussed in \cite{Sundman_JSAN2014,Rakotomamonjy_SP2011},  our discussion is more comprehensive with respect to  recent developments focusing on centralized as well as decentralized settings and complexity analysis. Moreover, \cite{Sundman_JSAN2014,Rakotomamonjy_SP2011} did not consider data gathering exploiting  spatio-temporal sparsity, design of data acquisition and reconstruction schemes to meet communication constraints   and the use of CS to solve any other inference tasks.}
    The review papers  \cite{Sun_TWC13,Sharma_COMT16,Salahdine_J16} have focused  mainly on cognitive radio networks, while some of the algorithms discussed are applicable for data gathering in sensor network applications as well. {To the best of the authors' knowledge, there is no any review paper that discusses CS based inference or impact of practical aspects on CS based processing in WSNs.  }  A summary of survey/review papers most related to this paper is given in Table \ref{table_relatedpapers}.

% Most of the available review/survey papers related to sensor networks while there are several papers specifically focusing in sensor networks. Relevant to this paper,  the applications of CS in  communication networks have been  reviewed  in  \cite{Sun_TWC13,Sharma_COMT16,Salahdine_J16,Qaisar_13,Choi_COMT2017,Pudlewski_COMT2013,
%Huang_axv13,Balouchestani_MWN2011,Choi_COMT2017,Sundman_JSAN2014}.   {\bf The applications of CS in WSNs have been discussed under the general topic of communication systems \cite{Qaisar_13}. }
%
%CS in wireless communication networks \cite{Qaisar_13,Choi_COMT2017}.
% There are some papers on applications of CS in wireless communication in general of which  WSNs can be considered to be a special case. Most of the survey papers on applications of CS in general wireless communication systems  have focused  on several physical layer operations \cite{Qaisar_13,Huang_axv13,Choi_COMT2017}.

%\begin{figure*}[t]
%\centering
%\includegraphics[width=1.0\textwidth]{col1.pdf}
%\caption{Test}
%\end{figure*}

\begin{figure*}
\centering
\includegraphics[width=5.0in]{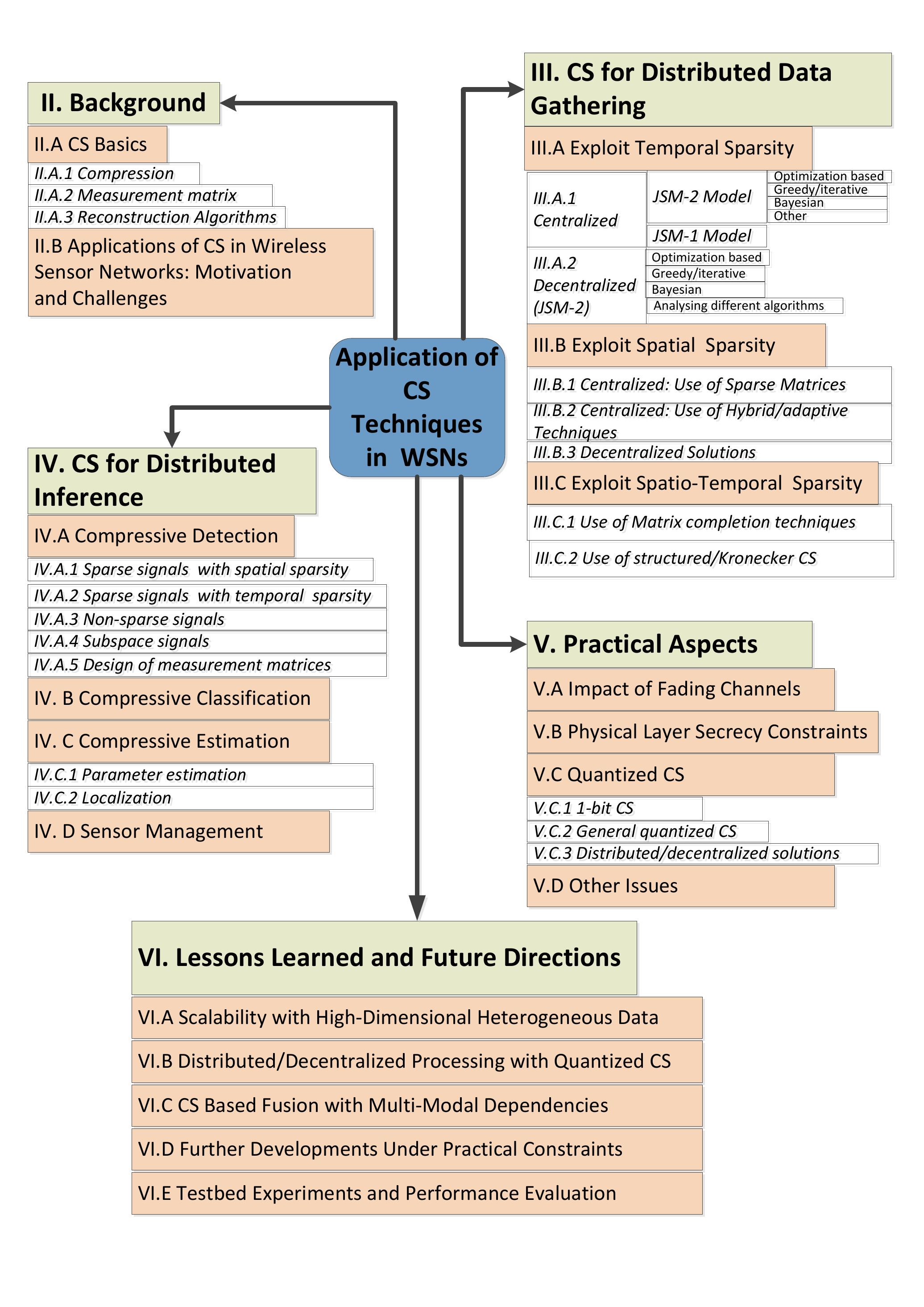}
 \caption{Roadmap of the paper}
\label{fig_road_map}
\end{figure*}
The organization of the paper  is summarized below.
\subsection{Paper Organization}
The roadmap of the paper is shown in Fig. \ref{fig_road_map}. In Section \ref{sec_background}, CS basics and motivation behind its use in several WSN applications are discussed. Application of CS in efficient data gathering exploiting temporal and spatial sparsity is   reviewed in Section  \ref{sec_data_gathering}.  Furthermore,  data gathering techniques and  algorithms developed in centralized as well as in distributed/decentralized settings are discussed.  CS based inference including detection, classification and localization  is reviewed in Section \ref{sec_inference}. In Section \ref{sec_practicalCon}, a review on CS based signal processing under practical communication  considerations such as channel fading, physical layer secrecy and quantization is given. Finally future research directions are discussed in Section \ref{sec_future} and concluding remarks are given in Section \ref{sec_conclusion}.

\subsection{Notation and Abbreviations}
 Throughout the paper, we use the following notation.   Scalars (in $\mathbb R$) are  denoted by lower case letters and symbols, e.g., $x$ and $\theta$.  Vectors (in $\mathbb R^N$) are written in boldface lower case letters and symbols, e.g., $\mathbf x$ and $\boldsymbol\beta$.  Matrices are written in boldface upper case letters or symbols, e.g., $\mathbf A$ and $\boldsymbol\Psi$. By $\mathbf 0$ and $\mathbf 1$, we denote the  vectors with appropriate dimension in which all elements are zeros and ones, respectively. The identity matrix with appropriate dimension is denoted by $\mathbf I$.  The  transpose of  matrix $\mathbf A$ is denoted by $\mathbf A^T$.  The $i$-th row vector  and the $j$-th column vector  of matrix $\mathbf A$ are denoted by $\mathbf a^{i}$ and $\mathbf a_{j}$, respectively. The $(i,j)$-the element of  matrix $\mathbf A$ is denoted by $\mathbf A_{i,j}$ or $\mathbf A[i,j]$. The $i$-th element of  vector $\mathbf x$ is denoted by $\mathbf x[i]$ or $x_i$.  The $l_p$ norm is denoted by  $||.||_p$ while $|.|$ is used for  both the cardinality (of a set) and the absolute value (of a scalar). The Frobenius norm of $\mathbf A$ is denoted by $||\mathbf A||_F$. The  Hadamard (element-wise) product is denoted by $\odot$ while the Kronecker product is denoted by $\otimes$.   The notation  $\mathbf x \sim \mathcal N (\boldsymbol\mu, \Sigma)$ means  that the random vector $\mathbf x$ is distributed as multivariate  Gaussian with mean $\boldsymbol\mu$ and covariance matrix $\Sigma$.
The abbreviations used in the paper are summarized in Table \ref{table_abbreviations}.
\begin{table*}[!t]
\renewcommand{\arraystretch}{1.3}
 %if using array.sty, it might be a good idea to tweak the value of
% \extrarowheight as needed to properly center the text within the cells
\caption{Abbreviations used throughout the paper}
\label{table_abbreviations}
\centering
% Some packages, such as MDW tools, offer better commands for making tables
% than the plain LaTeX2e tabular which is used here.
\begin{tabular}{|l|l|l|l|}
\hline
Abbreviation & Description& Abbreviation & Description\\
\hline
%\hline
ACIE & Alternating Common and Innovation Estimation &  IST & Iterative  Soft Thresholding \\
ADM  & Alternating Direction Method  & JSM & Joint Sparsity Model  \\
AFC & Analog Fountain Codes &  LASSO & Least Absolute Shrinkage and Selection Operator \\
ALM & Augmented Lagrangian Multiplier  & LMS & Least-Mean Squares  \\
AMP & Approximate Message Passing & MAC & Multiple Access Channel   \\
AWGN & Additive White Gaussian Noise& MAP  & Maximum \emph{a posteriori} \\
BCD & Block-Coordinate Descent &  ML & Maximum Likelihood\\
BCS & Bayesian  CS & MMV & Multiple Measurement Vector  \\
BIHT & Binary IHT& MSP & Matching Sign Pursuit  \\
BP & Basis Pursuit & MT-BCS & Multitask BCS  \\
BPDN & Basis Pursuit Denoising & NHTP & Normalized HTP  \\
BSC & Binary Symmetric Channel & NIHT & Normalized IHT  \\
BSBL & Block  SBL & NP & Neyman Pearson \\
CB-DIHT & Consensus Based Distributed IHT & OMP & Orthogonal Matching Pursuit\\
CB-DSBL & Consensus Based Distributed SBL & PoI & Phenomenon of Interest\\
CoSaMP & Compressive Sampling Matching Pursuit & RIP & Restricted Isometry Property \\
CS & Compressive Sensing& RLS & Recursive Least Squares \\
CRLB & Cram$\grave{e}$r-Rao Lower Bound & RSSI & Received Signal Strength Indicator \\
CWS & Compressive Wireless Sensing&  RVM & Relevance Vector Machines\\
DBS & Distributed BP & SBL & Sparse Bayesian Learning \\
DC-OMP & Distributed and Collaborative OMP & SCoSaMP & Simultaneous CoSaMP \\
DCSP & Distributed and Collaborative SP  & SCS & Sequential CS \\
DCT & Discrete Cosine Transform & SDP & SemiDefinite Programming \\
DIHT & Distributed IHT & SHTP & Simultaneous HTP\\
DiOMP & Distributed OMP & SIHT & Simultaneous  IHT   \\
DiSP & Distributed SP & SiOMP & Side information based OMP\\
DOA & Direction of Arrival & SMV & Single  Measurement Vector\\
DOI & Difference-of-Innovations & SNHTP & Simultaneous NHTP\\
DR-LASSO & Decentralized  Row-based LASSO & SNIHT & Simultaneous NIHT\\
DWT & Discrete Wavelet Transform & SNR & Signal-to-Noise Ratio\\
FOCUSS & FOCal Underdetermined System Solver & SOCP & Second-Order Cone Programming\\
FPC & Fixed Point Continuation  & S-OMP & Simultaneous OMP\\
GAMP & Generalized AMP & SP & Subspace Pursuit \\
GLRT & Generalized Likelihood Ratio Test& SR & Sparse  Representation  \\
GMM & Gaussian Mixture Model & SSP & Simultaneous SP\\
GMP & Greedy Matching Pursuit &  TDOA &  Time-difference-of-arrival  \\
GSM & Gaussian Scale Mixture & TECC & Transpose Estimation of Common Component \\
GSP & Generalized SP& VBEM & Variational
Bayesian expectation-maximization  \\
HTP & Hard Thresholding Pursuit & VQ & Vector Quantizers   \\
IHT & Iterative Hard Thresholding & WSN & Wireless Sensor Network   \\
IoT & Internet of Things  & $~$ & $~$ \\
%IoT & Internet of Things & $~$ & $~$ \\
%$~$ & $~$\\
\hline
\end{tabular}
\end{table*}

\section{Background}\label{sec_background}
{We start our discussion by} presenting  some  background material  on   CS along with  motivating factors behind  its application in WSNs.
\subsection{CS Basics}\label{sec_CS Basics}
Let $\mathbf x\in \mathbb R^N$ be a discrete time signal vector.  When represented in an appropriate basis $\boldsymbol\Phi\in \mathbb R^{N\times N}$ so that $\mathbf x = \boldsymbol\Phi\mathbf s$, $\mathbf x$  is said to be sparse (with respect to the basis $\boldsymbol\Phi$) if $\mathbf s$ contains only a few nonzero elements; i.e., $||\mathbf s||_0 \ll N$. The support of $\mathbf s$ (also known as the sparsity pattern/sparse support set) is defined  as the set $\mathcal U\in \{1,\cdots,N\}$ such that
\begin{eqnarray*}
\mathcal U := \{i\in \{1,\cdots,N\}~|~ \mathbf s[i] \neq 0 \}
\end{eqnarray*}
where $\mathbf s[i]$ denotes the $i$-th element of $\mathbf s$ for $i=1,\cdots, N$.
\subsubsection{Compression}
In the  CS framework, compression of $\mathbf x$ is performed  using the following linear operation:
 \begin{eqnarray}
 \mathbf y = \mathbf A \mathbf x \label{smv_model_nonoise}
 \end{eqnarray}
 where $\mathbf A$ is a $M\times N$ linear projection (measurement) matrix with $M <N$. In the presence of noise, \eqref{smv_model_nonoise} can be represented in a more general form,
\begin{eqnarray}
\mathbf y =\mathbf A \mathbf x + \mathbf v \label{smv_model_noise}
\end{eqnarray}
where $\mathbf v$ denotes the $M\times 1$  additive  noise vector.
With a SMV, one aims to solve  for sparse $\mathbf s$ (equivalently $\mathbf x$ since $\mathbf x = \boldsymbol\Phi\mathbf s$ with known $\boldsymbol\Phi$) from  \eqref{smv_model_nonoise} or \eqref{smv_model_noise}.

%\subsubsection{Signal reconstruction }
Recovering $\mathbf s$ from its  compressed form $\mathbf y$ in (\ref{smv_model_nonoise}) (or \eqref{smv_model_noise}) is  ill-conditioned when $M < N$, however,  it has been shown that it  is possible to reconstruct $\mathbf s$
 under certain
conditions on the measurement matrix $\mathbf A$ if $\mathbf s$ is sufficiently sparse \cite{candesTao_TIT06,candes_TIT06,donoho_TIT06}. Reconstruction of $\mathbf s$ is exact when there is no noise and approximate when there is noise.

\subsubsection{Requirements for the measurement matrix}
Several matrix properties have been discussed  to establish necessary and sufficient conditions satisfied by the matrix $\mathbf A$ so that $\mathbf s$ can be recovered from $\mathbf y$ \cite{Eldar_B1,candesTao_TIT06,candes_TIT06,donoho_TIT06}. One  such property  is the \emph{restricted isometry property (RIP)} property. The matrix $\mathbf A$ is said to satisfy  RIP of order $k$ if  there exists a $\delta_k\in (0,1)$ such  that
\begin{eqnarray}
(1 - \delta_k)||\mathbf x||_2^2 \leq ||\mathbf A \mathbf x||_2^2 \leq (1 + \delta_k)||\mathbf x||_2^2
\end{eqnarray}
for all $\mathbf x \in \Sigma_k$ where $\Sigma_k= \{\mathbf x: ||\mathbf x||_0 \leq k\}$.  It has been shown that when the entries of $\mathbf A$ are chosen according to a Gaussian (mean $0$ and variance $\frac{1}{M}$), Bernoulli ($+\frac{1}{\sqrt{M}}$ or  $-\frac{1}{\sqrt{M}}$ with equal probability) or in general from a sub-Gaussian distribution, $\mathbf A$ satisfies RIP with high probability when $M=\mathcal O(k \log (N/k))$ \cite{Eldar_B1}.

%While random matrices have desirable properties, their implementation can be challenging in practice. other structured matrices have been investigated in \cite{Rauhut_bookc2010}

\subsubsection{Reconstruction algorithms}
In order to recover $\mathbf s$ from $\mathbf y$, the natural choice is to solve the following optimization problem (with no noise) \cite{Eldar_B1,candesTao_TIT06,candes_TIT06,donoho_TIT06}:
    \begin{eqnarray}
    \underset{\mathbf s}{\min} ||\mathbf s||_0 ~ \mathrm{such}~ \mathrm{that}~ \mathbf y = \mathbf A \boldsymbol\Phi \mathbf s.\label{l_0_min}
    \end{eqnarray}
    Unfortunately, this $l_0$ norm minimization problem is generally computationally intractable. In order to approximately solve (\ref{l_0_min}),   several  approaches have been proposed. One of the commonly used approaches  is to replace the  $l_0$ norm in (\ref{l_0_min}) with a convex $l_1$ norm. Greedy pursuit and iterative algorithms are also promising in approximately solving (\ref{l_0_min}). In the following, we briefly discuss these approaches.

\begin{itemize}
\item Convex relaxation:  A fundamental  approach for signal reconstruction  proposed in  CS theory is the so-called  basis pursuit (BP) \cite{Chen_SIAM98} in which the $l_0$ term in \eqref{l_0_min} is replaced by the $l_1$ norm to get
      \begin{eqnarray}
    \underset{\mathbf s}{\min} ||\mathbf s||_1 ~ \mathrm{such}~ \mathrm{that}~ \mathbf y = \mathbf A \boldsymbol\Phi \mathbf s.\label{l_1_relax}
    \end{eqnarray}
    Under some favorable conditions, the solution to (\ref{l_0_min}) coincides with that in (\ref{l_1_relax}). In the presence of noise,  basis pursuit denoising (BPDN) \cite{Chen_SIAM98}  aims at solving
    \begin{eqnarray}
      \underset{\mathbf s}{\min} ||\mathbf s||_1 ~\mathrm{such} ~\mathrm{that} ~ ||\mathbf y - \mathbf A \boldsymbol\Phi\mathbf s||_2 \leq \epsilon_1 \label{BPDN}
     \end{eqnarray}
            and least absolute shrinkage and selection operator (LASSO)   \cite{Tibshirani_JoRoy1996,Beck_SIAM2009,Bickel_Ann2009}  solves
     \begin{eqnarray}
      \underset{\mathbf s}{\min}  ||\mathbf y - \mathbf A \boldsymbol\Phi\mathbf s||_2 ~\mathrm{such} ~  \mathrm{that} ~||\mathbf s||_1 \leq  \epsilon_2\label{Lasso}
     \end{eqnarray}
     or equivalently
      \begin{eqnarray}
      \underset{\mathbf s}{\min}  \lambda ||\mathbf s||_1 + \frac{1}{2}||\mathbf y - \mathbf A \boldsymbol\Phi\mathbf s||_2 \label{Lasso_2}
     \end{eqnarray}
     where $\lambda$ is the penalty parameter and  $\epsilon_1, \epsilon_2  > 0$. In order to further enhance the performance of the $l_1$ norm minimization based approach, the authors in \cite{Candes_JFAA2008} have proposed to optimize the reweighted $l_1$ norm. The reweighted $l_1$ norm form of LASSO in \eqref{Lasso_2} reduces to
      \begin{eqnarray}
      \underset{\mathbf s}{\min}  \sum_{i=1}^N w_i |\mathbf s[i]| + \frac{1}{2}||\mathbf y - \mathbf A \boldsymbol\Phi\mathbf s||_2 \label{Lasso_2_rew}
     \end{eqnarray}
     where $w_i > 0$ denotes the weight  at  index $i$.
     While convex optimization based techniques  are promising in providing optimal and/or near optimal solutions to  the sparse signal recovery problem, their computational complexity is relatively high. For example, the computational complexity of BP when interior point method is used scales as $\mathcal O(M^2 N^{3/2})$ \cite{Dai_TIT2009}.
     %There are recently proposed -- ADMM.  with a certain loss of recovery performance  is desirable in real-time applications.
     To reduce  computational complexity of sparse signal recovery,  greedy and iterative algorithms as discussed below have been proposed.

    \item Greedy and iterative algorithms: Greedy/iterative  algorithms aim to  solve \eqref{l_0_min} (or its noise resistant extension) in a greedy/iterative  manner which are in general less computationally complex than the optimization based approaches.  Examples of such algorithms include orthogonal matching pursuit (OMP) \cite{tropp_TIT2007}, subspace pursuit (SP) \cite{Dai_TIT2009}, Compressive sampling matching pursuit (CoSaMP)\cite{Needell_ACHN2009}, iterative hard thresholding (IHT) \cite{Blumensath_JFAA2008,Blumensath_ACHN2009} and their variants such as regularized OMP \cite{Needell_FCM2009}, and stagewise OMP \cite{donoho_TIT2012},  Normalized IHT (NIHT) \cite{Blumensath_JSTSP2010} and Hard threshoding pursuit (HTP) \cite{Foucart_SIAM2011}. The OMP and SP algorithms can be implemented with a computational complexity in the order of $\mathcal O(kMN)$ \cite{Dai_TIT2009,Qaisar_13} while the complexities  of  CoSaMP and IHT scale as $\mathcal O(MN)$  \cite{Qaisar_13} and $\mathcal O(MNT_r)$ \cite{Blumensath_JFAA2008,Blumensath_ACHN2009}, respectively where $T_r$ denotes the number of iterations required by the IHT algorithm for convergence. The recovery performance of OMP and IHT  is comparable  to the $l_1$ norm minimization based approach when the signal is sufficiently sparse ($k$ is very small) and the  SNR is relatively large \cite{tropp_TIT2007,Dai_TIT2009,Blumensath_JFAA2008}. On the other hand,   SP can perform comparable to the  $l_1$ norm minimization based approach even with relatively large $k$ depending on the distribution of non-zero coefficients of the sparse signal.

           \item Bayesian algorithms: Another class of sparse recovery algorithms falls under  the Bayesian formulation. In the Bayesian framework, the sparse signal reconstruction problem is formulated as a random signal estimation problem after imposing a sparsity promoting probability density function (pdf) on $\mathbf x$ in \eqref{smv_model_noise}. A widely used sparsity promoting  pdf  is the Laplace pdf  \cite{Ji_TSP2008}. With Laplace prior,  $\mathbf x$ is imposed with the pdf,
       \begin{eqnarray}
       p(\mathbf x| \rho) = \left(\frac{\rho}{2}\right)^{\rho/2} e^{-\sum_{i=1}^N |\mathbf x[i]|}\label{Laplace_prior}
       \end{eqnarray}
       where $\rho > 0$.
When the noise $\mathbf v$ in \eqref{smv_model_noise} is modeled as  Gaussian with  mean $\mathbf 0$ and  covariance matrix  $\sigma_v^2 \mathbf I$, the solution in  \eqref{Lasso_2} corresponds to a maximum \emph{a posteriori} (MAP) estimate of  $\mathbf x$ with the prior \eqref{Laplace_prior}. Computation of the MAP estimator in  closed-form with the Laplace prior is computationally intractable, and several  computationally tractable algorithms have been proposed in the literature. Sparse signal recovery using sparse Bayesian learning (SBL)
 algorithms has been   proposed in \cite{Wipf_TSP2004}. In  \cite{Ji_TSP2008}, a Bayesian CS
framework has been  proposed where relevance vector machines (RVM) \cite{Tipping_JML2001} are used for signal
estimation after introducing a hierarchical prior which shares similar properties as the Laplace prior, yet, providing tractable computation. Babacan et.al.  in \cite{Babacan_TIP2010}  have also considered a hierarchical form of the Laplace prior of  which RVM is a special case.   CS via belief propagation has been  considered in \cite{Baron_TSP2010}. Bayesian CS by means of expectation propagation has been   considered in \cite{Seeger_ICML2008}. An interesting characteristic of the  Bayesian formulation is that it lets one  exploit  the statistical dependencies of the signal or dictionary  atoms while  developing sparse signal recovery algorithms. The authors  in \cite{Peleg_TSP2012} have considered the problem of sparse signal recovery in  the presence of correlated dictionary atoms in a Bayesian framework. While Bayesian approaches provide more flexibility in designing recovery algorithms than deterministic approaches  their computational cost is relatively high compared to greedy and  iterative techniques. For example,  the computational complexity of SBL scales as  $\mathcal O(N^3)$ \cite{Choi_COMT2017}.
\end{itemize}
Comparisons  of different sparse signal recovery algorithms in terms of  computational complexities and the minimum number of measurements needed can be found in  \cite{Qaisar_13,Zhang_Access15}.

In the following section, we discuss the motivation behind applying CS techniques  in WSN applications.

\subsection{Applications of CS in Wireless Sensor Networks: Motivation and Challenges}\label{sec_motivation_WSN}
In a WSN deployed to collect field information in different application scenarios such as environment monitoring and surveillance, gathering sensed information at distributed sensors  in
an energy efficient  manner is critical to the operation of  the sensor network for a long period of time. Since the   energy consumption of distributed sensors is mostly dominated by the radio  communication  \cite{Razzaque_Sen2014} and sensing \cite{Cheng_Globecom2010}, data compression prior to transmission is vital.  Further, the data collected at multiple nodes can be redundant in temporal or spatial domains, thus transmitting raw data may be  inefficient.
Data compression in WSNs  has been studied for  a long time focusing on redundancy in temporal and/or spatial domains.  A nice review on different compression schemes proposed for WSNs can be found  in \cite{Razzaque_Sen2014}.  Most of the existing schemes can be categorized as transform  coding \cite{Razzaque_Sen2014,haupt_SPM2008} and  distributed source coding \cite{Cristescu_CCC2004,Baron_2006} which basically  suffer from the requirement of in-network computation and control overheads.  To that end, the CS framework, as a  successful  way of data compression not only for data gathering but also for solving other inference tasks,  has been found to be attractive   in the recent years. CS based compression does not require intensive computation at sensor nodes and complicated transmission control compared to conventional compression techniques.  In order to further reduce the computation and communication costs at the local nodes, quantized CS \cite{Boufounos_CISS2008,Gupta_ISIT2010,Boufounos_ICASSP2010,
Jacques_TIT13,Boufounos_Asilomar09} can be utilized.

As such, an immediate  application of CS in WSNs is compressive data gathering. When merging CS and data gathering, one of the main issues to be considered is how to design the two parts;  compression side and the reconstruction side,  under communication constraints. In WSNs, data collected at multiple nodes may have low dimensional properties in different dimensions; e.g., spatial, temporal,  or spatio-temporal.  On the compression  side, practical  design of compression schemes via random projections depends on how the sparsity is exploited.   CS theory was initially developed for estimating a single sparse signal using   SMV  \cite{candes_TIT06,candes_SPM08,donoho_TIT06,candesTao_TIT06}, and then it was extended to estimate multiple sparse signals using  MMVs  \cite{Tropp3,Tropp4,Chen_TSP2006} under certain assumptions.  While the CS reconstruction techniques developed for the  SMV and  MMV cases can be applied for CS based data gathering in ideal situations, extensions to the standard CS framework was required to account for communication related aspects.  In early works on  applying CS based techniques for data gathering, it was mainly assumed that the sensor nodes communicate with a  fusion center, and signal reconstruction is performed at the fusion center \cite{Luo_mobicom2009}.  In this approach,  multi-hop communication between the sensors and the fusion center incurs a significant communication cost. In order to reduce the communication overhead, one of the approaches explored widely is to design sparse, structured or adaptive measurement matrices \cite{wang_IPSN07} which can be different from  'good' CS measurement matrices. When using such matrices for compression, it is  required to establish the conditions under which the successful recovery is guaranteed. Another approach is to minimize the communication overhead is to perform reconstruction in a decentralized manner which requires sensor nodes to communicate only with their  one-hop neighbors. Decentralized processing is attractive in WSN applications since it improves the scalability and robustness compared to centralized processing.  This approach requires the extension of the CS recovery algorithms into the decentralized setting.

In certain WSN applications where compression via CS seems to be promising, complete signal reconstruction as required for data gathering is not required. For example, in detection,  classification and parameter estimation problems, it is more important to understand the amount of information retained in the compressed domain so that a reliable inference decision  can be obtained without complete signal reconstruction. In these inference problems  with
compressed data, the performance and the specific design principles depend on how
the signals or the parameters are modeled.  Further, the robustness and the accuracy of the CS framework under other practical communication considerations such as  fading channels, secrecy issues, measurement  quantization, link failures and missing data  need to be understood to make CS based techniques feasible in WSN applications.
{In the recent literature, the CS framework has been extended to cope with a variety of practical aspects. }
%In particular, the applications of CS in WSNs have stimulated a variety of new problems that require the  extension of  the CS framework.
The rest of the sections  are devoted to a discussion of   recent efforts of such  extensions in some detail.

\section{CS for Distributed  Data Gathering  }\label{sec_data_gathering}
In this section, we discuss  the CS based data gathering problem formulated as a sparse signal reconstruction problem  exploiting temporal,  spatial and spatio-temporal sparsity in WSNs. We describe the modifications and extensions to the CS framework to take communication constraints  into account focusing on  compression side as well as the recovery algorithms considering centralized and  distributed/decentralized settings.

\subsection{CS Based Data Gathering Exploiting  Temporal Sparsity}\label{sec_temporal}
\begin{figure}[!t]\centering
\includegraphics[width=2.0in]{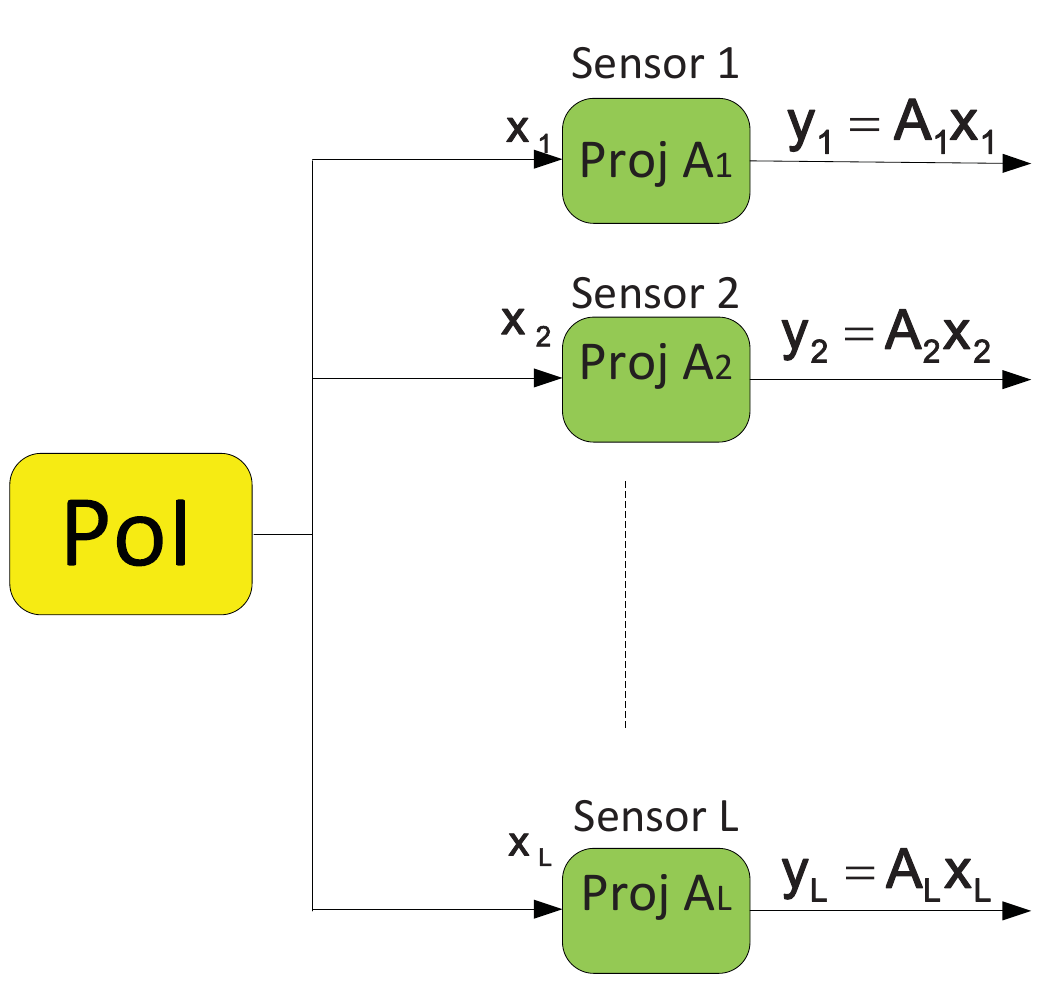}
 \caption{Acquisition of compressed measurements of  observations with temporal sparsity}
\label{fig_temporal_sparsity}
\end{figure}

Consider a WSN with $L$ sensor nodes observing a PoI as illustrated in Fig. \ref{fig_temporal_sparsity}. The time samples  collected at the $j$-th node are  represented  by the vector  $\mathbf x_j\in \mathbb R^N$. In data gathering, sensor readings need be collected efficiently. Transmitting raw data vectors $\mathbf x_j$'s is inefficient in many applications since sensor data is compressible   with many types of  sensors. Data collected at acoustic, seismic, IR, pressure, temperature, etc., can have sparse representation in a certain basis. For example,  audio signals collected by acoustic sensors (microphones) can be sparsely represented in DCT and DWT \cite{Griffin_EUSIPCO2008}. Formally,
with  an orthonormal basis $\boldsymbol\Psi_j\in \mathbb R^{N\times N}$, when $\mathbf x_j$  is represented as $\mathbf x_j = \boldsymbol\Psi_j \mathbf s_j$ for $j=1,\cdots,L$ , $\mathbf x_j$ is said to be sparse when $\mathbf s_j$ contains only a few nonzero elements compared to $N$. Thus, CS is readily applicable to compress  temporal sparse data. In particular, only a small number of random projections obtained via $\mathbf y_j=\mathbf A_j \mathbf x_j$ at the $j$-th node for $j=1,\cdots,L$ (Fig. \ref{fig_temporal_sparsity})   are sufficient to be transmitted where  $\mathbf A_j \in \mathbb R^{M\times N}$, $M < N$ is the measurement  matrix used at the $j$-th node.
When applying CS to compress temporal sparse data at a given node,  the  data streams collected at the node are compressed via random projections independently. The goal is to reconstruct $[\mathbf x_1, \cdots, \mathbf x_L]$ based on their compressed versions communicated through the network where the reconstruction techniques depend on  the specific communication architecture used to combine $\mathbf y_j$'s.

\subsubsection{Centralized solutions for simultaneous sparse signal recovery exploiting temporal sparsity}\label{sub_Sec_MMV_cent}
First, we focus on recovering $\mathbf X\equiv [\mathbf x_1, \cdots, \mathbf x_L]$ jointly based on $\mathbf Y = [\mathbf y_1, \cdots, \mathbf y_L]$ in a centralized setting. In this setting,   the  nodes transmit their  compressed measurements to a central fusion center with long-range single-hop communication.    The MMVs collected  at the fusion center in matrix form with the same projection matrix so that $\mathbf A_j = \mathbf A$ for $j=1,\cdots,L$ can be represented by
\begin{eqnarray}
\mathbf Y = \mathbf A \mathbf X\label{MMV_common}.
\end{eqnarray}
For the more general case with    different projection matrices, the observation matrix at the fusion center  can be expressed as
\begin{eqnarray}
\mathbf Y = [\mathbf A_1 \mathbf x_1, \cdots, \mathbf A_L \mathbf x_L].\label{MMV_different}
\end{eqnarray}
In the simultaneous sparse approximation framework,  $\mathbf X$  needs to be reconstructed when   $\mathbf Y $ (or its noisy version) and  $\mathbf A$ (with the same projection matrix) or $\mathbf A_1, \cdots, \mathbf A_L$ (with different matrices) are given.   In order to jointly estimate $\mathbf X$,  joint sparse structures of $\mathbf X$ can be exploited. There are several such structures discussed with applications to sensor networks \cite{Wakin_NIPS2005,Duarte_TIT13,Baron_2006}.  It is worth noting that estimation of $\mathbf X$ sometimes refers to estimating the support of $\mathbf X$ by some authors. While the techniques and performance guarantees are not the same for estimating $\mathbf X$ and its support, sometimes it is sufficient to estimate only the support jointly. This is because,  once the support set is known, estimating individual coefficients reduces to a linear estimation problem.

\subsubsection*{Common support set model (JSM-2)}
The widely considered joint sparse model for sensor network data  is the common support set  model  which is termed as the  JSM-2 model in \cite{Baron_2006}. In this model,  the sparse signals observed at multiple nodes, $\mathbf x_j$'s, have the same but unknown sparsity pattern with respect to the same basis. However, the corresponding amplitudes can be  different in general.  The JSM-2 model with  the same measurement matrix as in (\ref{MMV_common}) is commonly termed  as the MMV model \cite{Cotter_05,Chen_TSP2006}.  Without loss of generality, in the rest of the section, we assume that $\mathbf x_j$'s are sparse in the standard canonical basis unless otherwise specified.

While  developing algorithms and evaluating performance  with  the JSM-2 model, several measures have been defined.
To measure the number of nonzero elements of $\mathbf X$, $||\cdot||_{\mathrm{row}-0}$ norm is widely used where
\begin{eqnarray}
||\mathbf X||_{\mathrm{row}-0}= |\mathrm{rowsupp}(\mathbf X)|\label{l0}
\end{eqnarray}
with
\begin{eqnarray}
\mathrm{rowsupp}(\mathbf X) = \{i\in [1,\cdots,N]: \mathbf X_{i,j}\neq 0~ \mathrm{for} ~\mathrm{some} ~j\}.
\end{eqnarray}
The natural approach to solve for sparse $\mathbf X$ from  $\mathbf Y$  in (\ref{MMV_common})  is to solve the following optimization problem:
\begin{eqnarray}
\underset{\mathbf X}{\min} ~ ||\mathbf X||_{\mathrm{row}-0} ~\mathrm{such}~ \mathrm{that} ~ \mathbf Y = \mathbf A \mathbf X\label{l0_MMV}
\end{eqnarray}
with no noise or
\begin{eqnarray}
\underset{\mathbf X}{\min}  ||\mathbf X||_{\mathrm{row}-0} ~\mathrm{such}~ \mathrm{that}  \frac{1}{2}||\mathbf Y - \mathbf A \mathbf X||_F \leq \epsilon \label{l0_MMV_noise}
\end{eqnarray}
in the presence of noise.   Since the problem \eqref{l0_MMV} (and \eqref{l0_MMV_noise}) is NP hard, one often solves   \eqref{l0_MMV} (and \eqref{l0_MMV_noise}) by using the mixed norm approach  which is  an extension of the convex relaxation method for the  SMV case to the MMV case.

\subsubsection*{Convex relaxation}
A large class of relaxation versions of $||\mathbf X||_{\mathrm{row}-0}$ aims at  solving  an  optimization problem of the following form \cite{Rakotomamonjy_SP2011}:
\begin{eqnarray}
\underset{\mathbf X}{\min}  J_{p,q}(\mathbf X) ~\mathrm{such}~ \mathrm{that} ~ \mathbf Y = \mathbf A \mathbf X\label{mixed_norm_Nonoise}
\end{eqnarray}
in the noiseless case and
\begin{eqnarray}
\underset{\mathbf X}{\min} ~ \frac{1}{2} ||\mathbf Y - \mathbf A \mathbf X||_F^2 + \lambda  J_{p,q}(\mathbf X)\label{mixed_norm_noise}
\end{eqnarray} in the noisy case (also known as R-LASSO \cite{ling_icassp11}) where $\lambda$ is a penalty parameter
and
\begin{eqnarray}
J_{p,q}(\mathbf X) = \underset{i}{\sum} (||\mathbf x^{i}||_p)^q\label{J_pq}
 \end{eqnarray}
 where typically $p\geq 1$ and $q\leq 1$. It is noted that the value $q$  promotes the common sparsity profile of $\mathbf X$, $p$ is a measure to weight the contribution of multiple signals to the common sparsity profile and  \eqref{J_pq} is convex whenever $q=1$ and  $p\geq 1$.
 Different approaches have been discussed  to solve (\ref{mixed_norm_Nonoise}) and \eqref{mixed_norm_noise} for different values of $p$ and $q$.
 The most widely considered scenario is  $p=2$ and $q=1$ which  is commonly known as M-BP  \cite{Cotter_05,Malioutov_TSP2005,Stojnic_TSP2009,Eldar_TIT2010,Zhang_Access15,Lu_JMLR2011,
 Rakotomamonjy_SP2011,Yuan_Royal2006,Qin_MPC2010}. This case is also  dubbed as the mixed $l_2/l_1$ norm minimization approach.  When $L=1$, this case reduces to the BP (or LASSO) formulation in \eqref{l_1_relax} (or \eqref{Lasso_2}).
 The works in  \cite{Cotter_05,Malioutov_TSP2005,Stojnic_TSP2009,Eldar_TIT2010,Zhang_Access15,Lu_JMLR2011,
 Rakotomamonjy_SP2011,Yuan_Royal2006,Qin_MPC2010} have focused  on developing different algorithms, and establishing recovery guarantees.  In \cite{Cotter_05}, M-FOCUSS (FOCal Underdetermined System
Solver) has been  developed for the noiseless case, and the regularized M-FOCUSS for the noisy case. M-FOCUSS is an iterative algorithm that uses  Lagrange multipliers and is also applicable when  $q\leq 1$.
The average case analysis on recovery guarantees using multichannel BP  has been  discussed in \cite{Eldar_TIT2010}. In \cite{Lu_JMLR2011}, an  alternating direction methods (ADM) based approach has been  proposed to solve \eqref{mixed_norm_noise} which is called MMV-ADM.  The M-BP problem as a special case of group LASSO, with the   block coordinate descent algorithm called M-BCD,  has been  discussed in \cite{Yuan_Royal2006,Qin_MPC2010,Rakotomamonjy_SP2011}. In \cite{Stojnic_TSP2009}, the mixed $l_2/l_1$ type norm minimization problem has been  solved via  a semi-definite program which is shown  to  reduce to solving a second-order cone program (SOCP)-a special type of
semi-definite program.  In \cite{Tropp3}, the case where $p=\infty$ and $q=1$ has been  considered and the authors have  discussed the conditions under which the convex relaxation is capable of ensuring recovery guarantees.  Some known theoretical results on recovery guarantees of the SMV case have been generalized to the  MMV case in \cite{Chen_TSP2006} considering $q=1$ and arbitrary $p$.  Recovery guarantees using fast thresholded Landweber algorithms considering $p=1,2,\infty$, and $ q=1$ have been discussed in \cite{Fornasier_SIAM2008}.
A comparison of different simultaneous sparse approximation methods with different values for $p$ and $q$ can be found in \cite{Rakotomamonjy_SP2011,Tropp3}.

\begin{table*}
\renewcommand{\arraystretch}{1.3}
 %if using array.sty, it might be a good idea to tweak the value of
% \extrarowheight as needed to properly center the text within the cells
\caption{Centralized solutions  for joint sparse signal recovery with MMVs exploiting  temporal sparsity}
\label{table_temporal_sparse}
\centering
% Some packages, such as MDW tools, offer better commands for making tables
% than the plain LaTeX2e tabular which is used here.
\begin{tabular}{|l|l|}
\hline
Approach & Algorithms and References \\
\hline
\hline
{\bf JSM-2}\\
\hline
Convex relaxation, $p=2, q=1$ in \eqref{J_pq}  & Group LASSO (with BCD) \cite{Yuan_Royal2006,Qin_MPC2010,Rakotomamonjy_SP2011},  SDP-SOCP  \cite{Stojnic_TSP2009},  MMV-ADM  \cite{Lu_JMLR2011},  $\mathrm{MMV}_{\mathrm{prox}}$  \cite{Sun_NIPS2009} \\
\hline
Convex relaxation, $p=2, q\leq 1$ in \eqref{J_pq} & M-FOCUSS   \cite{Cotter_05}  \\
\hline
Convex relaxation, $p=1,2,\infty, q=1$ in \eqref{J_pq} & Landweber algorithms   \cite{Fornasier_SIAM2008}\\
\hline
Convex relaxation, $p=\infty, q=1$  in \eqref{J_pq}&    \cite{Tropp3} (Algorithm is implemented via standard mathematical software) \\
\hline
%Square of mixed $l_2/l_1$ norm   & \\
%\hline
Greedy and iterative & S-OMP   \cite{Tropp4},  SIHT, SNIHT \cite{Blanchard_TSP2014,Makhzani_icc12}, SHTP \cite{Foucart_SAMPTA11}, SNHTP \cite{Blanchard_TSP2014},  SCoSaMP \cite{Blanchard_TSP2014},  Generalized SP \cite{Feng_wcnc2013} \\
\hline
Bayesian & MSBL \cite{Wipf_TSP2007}, MMV-AMP \cite{Ziniel_JSTSP2013,Al-Shoukairi_Asilomar14}, T-MSBL \cite{Zhang_JTSP2011},   MT-BCS \cite{Ji_TSP2009,Wu_SPL2015},  M-BCS-GSM \cite{Tzagkarakis_ICASSP10}\\
\hline
Other approaches & Rank-aware algorithms \cite{Davies_TIT2012}, block sparse signal recovery based algorithms \cite{Eldar_TIT2009},  BSBL \cite{Zhang_TSP2013}\\
\hline
{\bf JSM-1}\\
\hline
Convex relaxation & weighted $l_1$ norm minimization \cite{Baron_2006}, $l_1$ norm minimization using ADM \cite{Matamoros_TSIPN2015} \\
\hline
Greedy & SiOMP \cite{Zhang_NIDC2010}\\
\hline
Greedy/optimization based &  DOI and Texas DOI \cite{Valsesia_Asilomar2013} (to estimate sparse  innovation components, greedy/optimization based \\
$~$ & recovery algorithms can be used, valid  for JSM-3 model as well)\\
\hline
{\bf JSM-3}\\
\hline
Greedy/optimization based & TECC and ACIE \cite{Baron_2006}  (to estimate sparse  innovation components, \\
$~$ & greedy/optimization based recovery algorithms can be used)\\
\hline
\end{tabular}
%\begin{enumerate}
%\item
%  \end{enumerate}
\end{table*}

\subsubsection*{Greedy algorithms}
In order  to solve \eqref{l0_MMV},   several greedy and iterative algorithms have been proposed which typically have less computational complexity than convex relaxation based approaches. In particular, most of these approaches are extensions of  their SMV counterparts. The extension of the  OMP algorithm to the MMV case  with the common support set model, S-OMP, has been  considered in \cite{Tropp4}. Performance analysis of the   S-OMP algorithm including  noise robustness  has been  presented  in a recent paper \cite{Determe_TSP2017}.  An MMV extension of the IHT algorithm, SIHT has been considered in  \cite{Blanchard_TSP2014,Makhzani_icc12}. Some variants of SIHT such as SNIHT, and extensions of other greedy algorithms such as HTP, NHTP, and CoSaMP  have  also been discussed in \cite{Blanchard_TSP2014,Foucart_SAMPTA11}. Generalized SP as an extension of the SP algorithm with MMVs  has been  considered in \cite{Feng_wcnc2013} which is shown to outperform the natural extension of SP,  i.e., simultaneous SP (S-SP).

\subsubsection*{Bayesian algorithms}
Solutions to  the MMV problem with the  JSM-2  model in the Bayesian setting have been  discussed in \cite{Wipf_TSP2007,Ji_TSP2009,Zhang_JTSP2011,Tzagkarakis_ICASSP10,
Ziniel_JSTSP2013,Zhang_TSP2013,Al-Shoukairi_Asilomar14,Wu_SPL2015}.  The MSBL algorithm,  which  is an extension of the SBL algorithm with SMV, has been  developed in \cite{Wipf_TSP2007}. Multitask BCS (MT-BCS)  has been  developed  in \cite{Ji_TSP2009} in which applications of multitask learning to solve the MMV problem in the Bayesian framework have been  discussed. In \cite{Wu_SPL2015}, the  authors have extended the MT-BCS framework to take the intra-task dependency into account.    In \cite{Tzagkarakis_ICASSP10}, the MMV problem in a Bayesian setting has been  considered focusing on DOA estimation where the  prior probability distribution is modeled with a Gaussian scale mixture
(GSM) model. The proposed approach is dubbed M-BCS-GSM.  Belief propagation based methods have been  proposed in \cite{Al-Shoukairi_Asilomar14,Ziniel_JSTSP2013} in which the approximate message passing  (AMP) framework is used to jointly estimate the sparse signals.  In \cite{Zhang_JTSP2011}, the MMV  problem has been  solved when the  nonzero elements of a given column  of $\mathbf X$ are correlated and the two Bayesian learning algorithms, called  T-SBL and T-MSBL have been   developed.

\subsubsection*{Other approaches}
The MMV problem with the JSM-2 model has been  treated as a block sparse signal recovery problem  after vectorizing  $\mathbf X$ to a block sparse signal in \cite{Eldar_TIT2009}. Then,  the algorithms developed for block sparse signal recovery with SMV can be used to solve the MMV problem.   In \cite{Zhang_TSP2013}, block SBL (BSBL)  algorithms, which take intra-block correlations into account,  have been developed where the  MMV problem is  treated as a block sparse signal  recovery problem. In \cite{Davies_TIT2012}, the authors have shown that most simultaneous sparse recovery algorithms discussed above  are  rank blind. They have proposed rank aware algorithms with mixed norm minimization as well as greedy  algorithms which show better performance than rank blind algorithms under certain conditions. In \cite{Caione_TIF2014}, real temperature, humidity and light data  have been used to validate  some algorithms developed under the JSM-2 model exploiting spatial and time correlations
among network data. An adaptive sparse sensing framework has been proposed in \cite{Chen_TWC15} which precisely
recovers  spatio-temporal physical fields by optimizing for  sparse sampling patterns.

\subsubsection*{Common support set  + innovation model (JSM-1)}
Another widely considered joint sparse support set  model with many WSN applications is the common support set + innovation model, which is termed as the  JSM-1 model in \cite{Baron_2006}. It is assumed that each sparse signal $\mathbf x_j$ can be expressed as $\mathbf x_j = \mathbf x_C + \tilde{\mathbf x}_j$ where $\mathbf x_C$ is a common (to all $\mathbf x_j$'s) component which is sparse and $\tilde{\mathbf x}_j$ is called the innovation component which is also sparse but different from  $\mathbf x_C$'s  sparsity pattern. Recovery of $\mathbf X$  with the JSM-1 model has been  discussed in \cite{Baron_2006}. In \cite{Baron_2006},   weighted $l_1$ norm minimization for jointly estimating $\mathbf X$ is discussed. Considering the signal vector at one node as side information, the authors in \cite{Valsesia_Asilomar2013} have developed the Difference-Of-Innovations (DOI) and Texas DOI algorithms  for simultaneous sparse signal approximation for  the JSM-1 model. In \cite{Matamoros_TSIPN2015}, centralized (as well as distributed) implementation of the ADM algorithm with MMV has been  presented.
In \cite{Zhang_NIDC2010}, side information based OMP (SiOMP) has been  proposed. In  SiOMP,  the distributed CS is performed for the JSM-1 model where the estimate at one node via OMP is considered to be the initial value for the estimate at the next node.

Although not as interesting  as  the JSM-1 and JSM-2 models, some works can be found for the JSM-3 model discussed in \cite{Baron_2006}. In the JSM-3 model,  the signal observed at each node is assumed to be composed of a nonsparse common component + sparse innovation component. Applications of this model in sensor network settings have been  discussed in \cite{Baron_2006}. Further, two algorithms  called  Transpose Estimation of Common Component (TECC) and Alternating Common and Innovation Estimation (ACIE) have been developed  in \cite{Baron_2006}. In these algorithms, the nonsparse common support is computed jointly and using that the sparse innovation components are computed by running sparsity aware algorithms.  The algorithms  developed in \cite{Valsesia_Asilomar2013} for the JSM-1 model are applicable to the JSM-3 model as well.

%\subsubsection{JSM-3 Model}
In Table \ref{table_temporal_sparse}, we summarize the simultaneous sparse signal recovery algorithms classifying under different categories as discussed above  with  MMVs  for different JSM models.

\subsubsection{Decentralized solutions for simultaneous sparse signal recovery exploiting temporal sparsity}
While centralized solutions are attractive in terms of the performance, their implementation can be prohibitive in resource constrained WSNs since the power costs for long-range transmission  can be still quite high.  Further, centralized solutions are not robust to node and link failures.  Decentralized solutions are attractive and are sometimes  necessary  in resource constrained  sensor networks in which  distributed nodes exchange (locally  processed) messages only with their neighbors. They can  significantly reduce   the overall communication power and bandwidth compared to centralized processing. In the decentralized setting, each node is required to obtain the centralized solution (or an approximation to it) by only communicating within one-hop neighbors.  Since the initial CS framework was developed for the SMV or the MMV case with centralized architecture, there was a need to extend it to the distributed and decentralized framework  to address the necessities arising out of WSN applications.

In developing decentralized algorithms exploiting temporal sparsity,   where the goal is to estimate a set of sparse signals using their compressed versions, the joint sparse structures play an important role. In fact, most of the decentralized solutions developed so far consider  the JSM-2 model for multiple sparse signals.  First, we review the decentralized extensions of the optimization based algorithms where the goal is to solve equations of the form  \eqref{mixed_norm_Nonoise} or \eqref{mixed_norm_noise} in  decentralized manner.

\subsubsection*{Optimization based approaches}
With the JSM-2 model, the matrix  $\mathbf X$ has only a small number of nonzero  rows.  In other words, when the common support set is estimated jointly, the individual estimates of  $\mathbf x_j$'s can be estimated by the $j$-th node. For the JSM-2 model with different measurement matrices at multiple nodes,  the row-based LASSO (R-LASSO) formulation  has been  extended to the decentralized setting  in \cite{ling_icassp11}. Recall that R-LASSO in \eqref{mixed_norm_noise} with different measurement matrices reduces to
%The R-LASSO solves for $X$ from \cite{ling_icassp11}
 \begin{eqnarray}
 \underset{\mathbf X}{\min}  \frac{1}{2}\sum_{l=1}^L ||\mathbf y_l - \mathbf A_l \mathbf x_l||_2^2 + L \lambda J_{p,q}^{1/p}(\mathbf X)\label{J_pq_diff}
 \end{eqnarray}
where  $ J_{p,q}(\mathbf X)$ is as defined in \eqref{J_pq}.
 In \cite{ling_icassp11},  a decentralized   implementation of \eqref{J_pq_diff} has been  proposed  with $p=2$ and $q=1$. In particular, the authors have reformulated \eqref{J_pq_diff} as a consensus optimization problem and developed an iterative algorithm. It is noted that consensus optimization is a powerful tool that can be utilized in designing decentralized
networked  systems with distributed nodes.
%The idea is to obtain an optimization variable
%common to the neighboring nodes by  letting  each node hold a local copy of the
%variable and imposing  consensus constraints on the local
%copies.
In  \cite{ling_icassp11}, the $l$-th node  estimates the common support set and the individual coefficient vectors by solving a local optimization problem via augmented Lagrangian multiplier  (ALM) method based on the information  received from its local neighborhood. The algorithm requires each node to communicate a length $N$ message at each iteration to its one-hop neighbors, thus the communication burden is proportional to the total number of iterations required and  $N$.
In \cite{Ling_TSP2013}, with the same problem model as in \cite{ling_icassp11},  a non-convex optimization approach  has been  discussed  extending the  reweighed $l_q$ norm minimization approach proposed in \cite{Candes_JFAA2008}  for simultaneous sparse signal recovery in a decentralized manner. Similar to \cite{ling_icassp11}, the authors have reformulated  the decentralized sparse signal recovery problem as a consensus optimization problem and only a length $N$ weight vector needs to be exchanged among neighboring nodes during each iteration. In \cite{Ling_TSP2013}, the ADM method has been  used  to solve the corresponding optimization problem at a given node.
A decentralized extension to the reweighted $l_1$ norm minimization approach has  also been  discussed in a recent  work in \cite{Fosson_TSP2016}, where the authors propose to solve the problem using the iterative  soft thresholding algorithm (IST).  This algorithm requires each node to transmit a length $N$ messages  in its local neighborhood at each iteration and the number of iterations  depends on the IST convergence rate.

A special case of the  JSM-2 model is  the case where all the nodes observe the same sparse signal (same support and same coefficients). We refer to this case as the  common signal model where there is only a single sparse signal to be estimated with MMVs. Decentralized algorithms for this model have been developed in \cite{Ramakrishnan_SSP2011,Zeng_JSTSP11,Mota_TSP2012,Bazerque_TSP10,Mateos_TSP10}.  It is worth noting that the work reported in \cite{Zeng_JSTSP11,Bazerque_TSP10} is application specific to spectrum sensing in cognitive radio networks, while the algorithms are  applicable for WSNs with MMVs. In \cite{Bazerque_TSP10,Mateos_TSP10}, the distributed LASSO (D-LASSO) algorithm has been  employed  to compute the common LASSO estimator by collaboration among nodes. In \cite{Zeng_JSTSP11}, a similar formulation and approach  has been  adopted  as in \cite{Bazerque_TSP10}, however, their algorithm has faster convergence rate via proper fusion.   In \cite{Ramakrishnan_SSP2011,Mota_TSP2012},  it is assumed that each node has access to only partial information on  random projections of the common  signal.  In \cite{Ramakrishnan_SSP2011}, a decentralized gossip based algorithm has been  developed to implement the $l_1$ norm minimization approach in a decentralized manner. Each node solves the optimization problem via the projected-gradient
approach by communicating only with one-hop neighbors.  Formulating  the $l_1$ norm minimization problem as a bound-constrained quadratic program, the gradient computation at each step of the coordinate descent algorithm is expressed as a sum of quantities computed at each node applying distributed consensus algorithm. In \cite{Mota_TSP2012}, distributed basis pursuit (DBS) has been  developed based on the ADM method.

There are a  few works that extend the optimization techniques considering the JSM-1 model  to the decentralized setting.   In \cite{Matamoros_TSIPN2015}, the authors have proposed a distributed version of the ADM algorithm (dubbed as DADMM)  to implement an equivalent version of \eqref{J_pq_diff} in a decentralized manner where communication is  only with one hop neighbours.

\begin{table*}
\renewcommand{\arraystretch}{1.3}
 %if using array.sty, it might be a good idea to tweak the value of
% \extrarowheight as needed to properly center the text within the cells
\caption{Decentralized estimation of  sparse signals with multiple measurements exploiting joint sparse structures; Different cases are for different joint structures of multiple sparse signals, Case 1: JSM-2 with different coefficients, Case 2:  JSM-2  with the  same coefficients (common signal model), Case 3: JSM-1}
\label{table_example_distributed}
\centering
% Some packages, such as MDW tools, offer better commands for making tables
% than the plain LaTeX2e tabular which is used here.
\begin{tabular}{|l|l|l|l|}
\hline
Algorithm/References & Applicability &   Commun.  complexity & Features/average computational complexity per node \\
 $~$ & $~$ &  ($\#$ of transmitted  & $~$ \\
  $~$ & $~$ & messages   per node ) & $~$\\
 %$~$ & $~$ &   among its one hop neighbors) & $~$\\
\hline
{\bf Optimization based}  \\
\hline
 DR-LASSO   \cite{ling_icassp11} & Case I & $\mathcal O(NT_1 I_t)$ & ALM  to solve the optimization problem / $\mathcal O(I_t(N^2MT_1+n_0NT_2))$\\
\hline
 reweighted $l_1$ norm min.   \cite{Fosson_TSP2016}& Case I & $\mathcal O(N I_t)$ &IST to solve the optimization problem/$\mathcal O(I_t(N^2M+N^2+NM))$\\
 \hline
 reweighted $l_q$ norm min.   \cite{Ling_TSP2013}& Case I &  $\mathcal O(N I_t)$ & Use non-convex sum-log-sum penalty, ADM to  solve the optimization \\
 $~$ & $~$ & $~$ & problem / $\mathcal O(I_t((N^2+M^3+NM^2)T_3 + n_0 N))$\\
\hline
  Distributed ADMM \cite{Matamoros_TSIPN2015} &  Case III & $\mathcal O(2N I_t)$ & ADM complexity\\
  \hline
  D-LASSO \cite{Bazerque_TSP10,Zeng_JSTSP11} & Case III & $\mathcal O(NI_t)$ & ADM complexity\\
\hline
Distributed BP \cite{Mota_TSP2012} & Case II & $\mathcal O(NI_t)$ or $\mathcal O(NT_3I_t)$ &  ADM complexity \\
\hline
 $l_1$ norm min.  \cite{Ramakrishnan_SSP2011} & Case II & $\mathcal O(NI_t)$ & Projected-gradient
approach/dominated by $\mathcal O((N^2n_0 + N)I_t)$\\
\hline
{\bf Greedy and iterative}\\
\hline
 DiOMP   \cite{Sundman_SP2014} & Case I/Case III & $\mathcal O(k^2)$ & $\mathcal O(k^c(T_4 MN))$ \\
\hline
 DC-OMP 1  \cite{Wimalajeewa_ICASSP13,Wimalajeewa_TSP2014}&  Case I &   $\mathcal O(I_t^{\dagger})$ & Fusion of estimated indices   at each iteration /$\mathcal O(I_t (MN + L))$  \\
\hline
 DC-OMP 2\cite{Wimalajeewa_TSP2014} & Case I& $\mathcal O(I_t^{\dagger}+N I_t)$ & Fusion of estimated indices as well as correlation coefficients  \\
 $~$ & $~$ & $~$& in the neighborhood   at each iteration / $\mathcal O(I_t (MN+n_0N+L))$\\
%$~$ & SiOMP \cite{Zhang_NIDC2010} & Case III &
\hline
 DiSP  \cite{Sundman_Icassp2012} & Case I/Case III & $\mathcal O(kI_t)$ & SP operations, fusion via majority rule/ $\mathcal O(I_t kMN)$ \\
\hline
  DCSP  \cite{Li_ICASSP14,Li_TSP2016} & Case I &  $\mathcal O(kI_t)$ & SP operations, fusion via majority rule/$\mathcal O(I_t(MN+n_0))$\\
\hline
  GDCSP  \cite{Li_ICASSP14,Li_TSP2016} & Case I &  $\mathcal O((k + N)I_t)$ & SP operations, fusion of indices, measurements/$\mathcal O(I_t(MN+n_0N))$ \\
\hline
 DIPP \cite{Sundman_TSP2016} & Case III &$\mathcal O(kI_t)$ & Modification of SP, \\
  $~$ & $~$ & $~$&  fusion via consensus, expansion/ $\mathcal O(I_t(NMT_5+kn_0))$ \\
%\hline
%DiT  \cite{Fosson_Icassp2014} &  Case I & $\mathcal O(2NI_t)$ & \\
\hline
 D-IHT \cite{Patterson_icassp13} &  Case II & $\mathcal O(N^{\dagger}I_t)$ & IHT operations/$\mathcal O(M_pNI_t)$ \\
\hline
%CB-IHT \cite{Patterson_TSP2014} &  Case II & $\mathcal O(-)$ & \\
%\hline
{\bf Bayesian}\\
\hline
 DCS-AMP  \cite{Makhzani_CAMSAP2013} & Case I &  $\mathcal O(MNI_t)$ & Bernoulli Gaussian signal prior is used/$\mathcal O(I_t(MN+n_0N))$\\
 \hline
%DB-DCS \cite{ Chen_TWC16} & Case III &  $\mathcal O(N-)$ & Gaussian prior  is used/dominated by $\mathcal O(I_t(N^3 + N^2M))$\\
%\hline
 CB-DSBL \cite{Khanna_TSIPN2017} &  Case I &  $\mathcal O(NT_3I_t)$ & $\mathcal O(I_t((N^2+M^3+NM^2+NT_3 n_0))$\\
\hline
\end{tabular}
\begin{enumerate}
\item \footnotesize $N$: dimension of the unknown sparse signals ,$M$: number of compressed measurements per node, $k$: number of nonzero elements of the sparse signals in JSM-2 (sparsity index), $k^c \leq k$: number of nonzero elements with commons support set in JSM-1,    $L$: number of total nodes in the network, $n_0$: average  number of on-hop neighbors per each node
\item $I_t$ denotes the number of iterations required for convergence which is a  algorithm dependent variable
    \item $T_1$ and $T_2$ denote the number of inner loop iterations required by  DR-LASSO   \cite{ling_icassp11}
        \item $T_3$: a variable to denote inner loop iterations in ADM
        \item $T_4$:  number of inner loop iterations in DiOMP
        \item $T_5$: number of inner loop iterations in DiPP
        \item In DC-OMP 1, DC-OMP 2,  DCSP, and GDCSP:  $I_t \leq k$
        \item $^{\dagger}$ denotes that  messages need to be  communicated globally via multi-hop
        communication
        \item $M_p\leq M$: number of partial measurements at each node
        \end{enumerate}
\end{table*}

\subsubsection*{Greedy/iterative  approaches}
Distributed and decentralized versions of greedy/iterative  algorithms can be found   in \cite{Sundman_Icassp2012,Sundman_JSAN2014,Fosson_Icassp2014,
Patterson_icassp13,Patterson_TSP2014,Zhang_NIDC2010,Wimalajeewa_ICASSP13,Wimalajeewa_TSP2014,Sundman_SP2014}. In \cite{Sundman_Icassp2012,Sundman_SP2014},  distributed SP (DiSP) and Distributed OMP (DiOMP)  have been  developed  which are applicable to both JSM-1   and  JSM-2  models.  In DiSP,  at each iteration an estimate to the support set is updated based on the local support set estimates of the neighboring nodes. In each iteration, it is required to  transmit the whole support set among the neighborhood. It is noted that, the number of iterations required for convergence depends on the network topology and is fairly close to the sparsity index.  In \cite{Sundman_TSP2016}, the distributed parallel pursuit (DIPP)  algorithm has been  proposed for  the JSM-1 model. In DIPP,  fusion is performed to improve the estimate of  the common support set via local communication, and the estimated support set is used as side information for complete signal reconstruction. Embedding fusion within OMP iterations, in \cite{Wimalajeewa_TSP2014}, the authors have proposed two decentralized versions of the OMP algorithm, called DC-OMP 1 and DC-OMP 2 for the JSM-2 model. In DC-OMP 1, an estimate of the support index is computed similar to the standard OMP and those computed  indices at all the nodes  are fused at each iterations to get an more accurate   set of indices for the support set. In DC-OMP 2, fusion is performed at two stages;  to improve the initial estimate  of the support set at each node, measurement fusion is done within the one-hop neighborhood. Similar to DC-OMP 1, index fusion is performed  to get a more accurate index set. Both DC-OMP 1 and DC-OMP -2 can terminate with  less than $k$ number of iterations and require multi-hop communication for the index fusion stage since global knowledge is required. In \cite{Li_ICASSP14,Li_TSP2016}, the distributed and collaborative subspace pursuit (DCSP) algorithm has been developed for the JSM-2 model. The ideas developed in \cite{Li_ICASSP14,Li_TSP2016} are similar to that in \cite{Wimalajeewa_TSP2014} with OMP, however, communication overhead requirements are slightly higher in \cite{Li_ICASSP14,Li_TSP2016} than that for DC-OMP 1 and DC-OMP 2.
%A distributed iterative algorithm, dubbed, DiT has been  developed  the  JSM-2 model  in \cite{Fosson_Icassp2014} which aims at minimizing a cost function accounts for a least squared residual term, a term to promote consensus in common support set estimation by local communication and a term to promote sparsity.
The algorithm requires each node to communicate with its neighbors twice in each iteration to update the common support set. Distributed IHT (D-IHT) and a consensus based distributed IHT named (CB-DIHT) have been   proposed in \cite{Patterson_icassp13,Patterson_TSP2014} for the common signal model.

\subsubsection*{Bayesian   approaches}
Distributed and decentralized Bayesian algorithms for sparse signal recovery have been proposed in \cite{Makhzani_CAMSAP2013,Chen_TWC16,Khanna_TSIPN2017}.  In \cite{Makhzani_CAMSAP2013}, an approximate message passing (AMP) based decentralized algorithm, AMP-DCS,  is developed to reconstruct a set of jointly sparse signals with JSM-2. The sparse signals are assumed to have sparsity inducing Bernoulli-Gaussian signal
prior.  The algorithm requires each node communicate $MN$ messages at each iteration. In \cite{Chen_TWC16} a decentralized Bayesian CS framework has been proposed for the JSM-1 model where applying variational Bayesian approximation, the common support component and the innovation component are  decoupled.  Formulating the decoupled reconstruction problem as a set of decentralized
problems with consensus constraints, each node estimates its innovation component independently and the  common component jointly via local communication.
%We dub this algorithm as DB-DCS in Table \ref{table_example_distributed}.
In \cite{Khanna_TSIPN2017}, a Consensus
Based Distributed Sparse Bayesian Learning (CB-DSBL) algorithm  has been  proposed. The authors also exploit ADM to solve the  consensus optimization problems in the sparse Bayesian learning framework  at  each node.

\subsubsection*{Analysing different types of  decentralized  solutions}
The above discussed decentralized algorithms are summarized in Table \ref{table_example_distributed}. The three cases in Table \ref{table_example_distributed} correspond to the JSM-2 model, common signal model and the JSM-1 model. {Selection of one algorithm over the other mainly depends on the desired performance requirements and tolerable   communication and computational complexities. In terms of computational and communication complexities, it is worth mentioning that almost all the optimization based techniques require a relatively   large number of iterations to converge. Thus, the  communication cost per  each node is relatively high.  Further, the computational cost at each node scales at least quadratic with respect to $N$ resulting in a relatively large computational cost when dealing with high dimensional signals.    Similarly, the decentralized versions of Bayesian algorithms also require a relatively high computational cost in processing high dimensional signals. On the other  hand, most of the greedy approaches require only few iterations for algorithm termination \cite{Wimalajeewa_TSP2014,Li_ICASSP14,Li_TSP2016,Sundman_Icassp2012,Sundman_SP2014} that are comparable to the  sparsity index $k$. This results in relatively small communication overhead compared to implementing optimization based algorithms in a decentralized manner.  In particular, DC-OMP 1 and DC-OMP 2 have a  faster convergence rate which require  less than $k$ number of iterations. Further, the in-node computational complexity of greed algorithms is much smaller (mostly linear with $N$) than that with optimization based and Bayesian approaches.   In terms of performance, optimization based  techniques with  a suitable penalty parameter performs better than the other types of algorithms   especially when $k$ is large and SNR is relatively  small. While both Bayesian and optimization based methods show similar order of computational complexity per node, Bayesian methods are more flexible in terms dependency on  parameters than the optimization based techniques. Bayesian CS techniques can be implemented fully automated since all the unknown parameters are estimated during the execution of the algorithm  while  the optimization based techniques require the tuning of parameters such as penalty parameter and     noise statistics for optimal performance. Thus, the optimization based techniques  may require a larger communication overhead than the Bayesian techniques since the communication overhead depends on the number of iterations of the algorithm. Moreover, the Bayesian approach provides a framework to account for the spatial and temporal statistical dependencies  of the joint sparse signals.  Comparing all types of algorithms, the greedy and iterative algorithms appear to be quite promising in decentralized processing of temporal sparse signals under severe  network resource constraints although a certain performance loss can be expected compared to the  optimization based and Bayesian approaches.}

\subsection{CS Based Data Gathering/Reconstruction Exploiting Spatial Sparsity}\label{sec_spatial}
\begin{figure}[!t]
\centering
\includegraphics[width=3.5in]{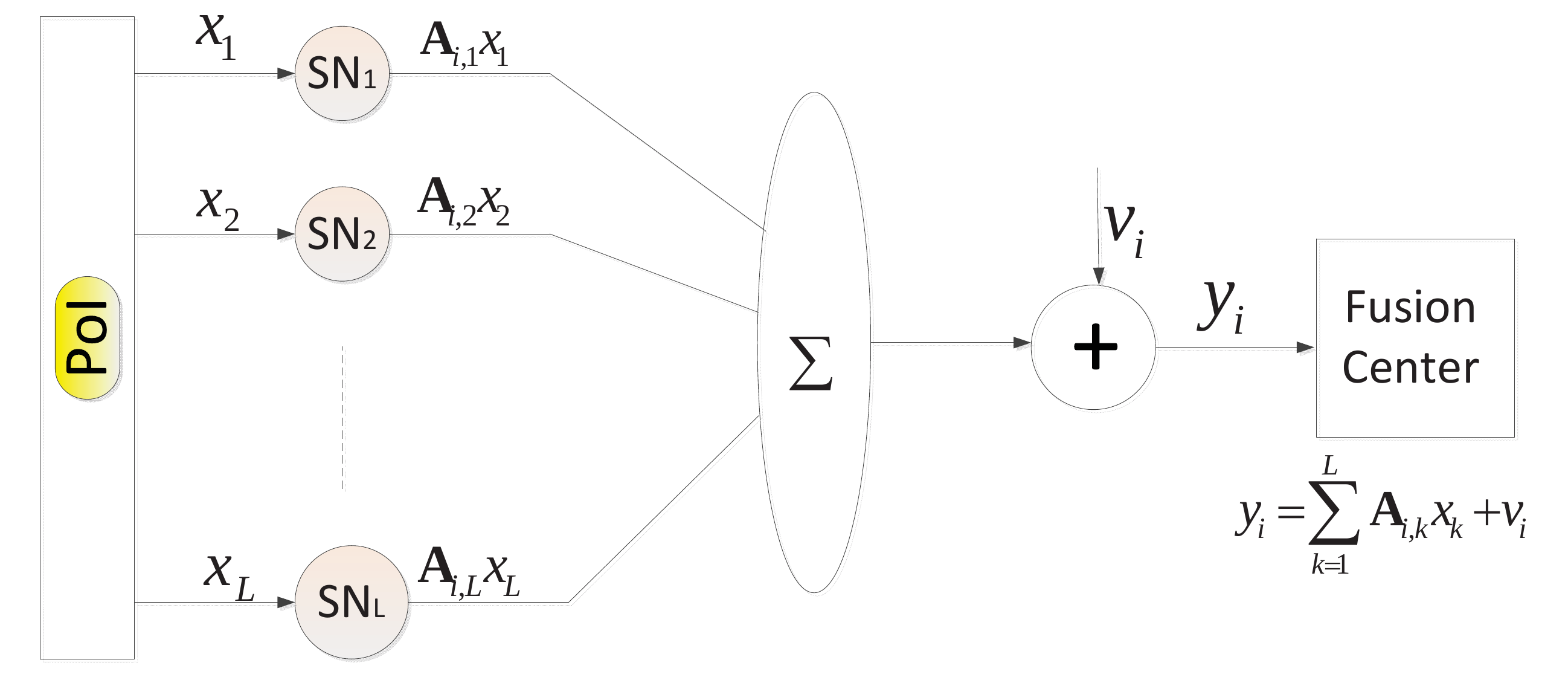}
 \caption{MAC model for CS based data gathering exploiting spatial sparsity \cite{haupt_SPM2008,Bajwa_IT2007}}
\label{fig_MAC}
\end{figure}

To exploit CS techniques for data gathering  exploiting spatial sparsity, compression of spatial data (over the nodes) via random projections  at a given time step is needed.  In contrast to  temporal data compression as considered in Section \ref{sec_temporal} where individual nodes use random projection matrices independently, in this case,  random projections have to be implemented in a distributed manner.
 Several architectures   have been explored  to implement distributed random projections so that a compressed version of sparse spatial data is made available at the fusion center.
One of the architectures,  with one-hop communication between the sensors and the fusion center is  known as  compressive wireless sensing (CWS) as proposed in \cite{haupt_SPM2008,Bajwa_IT2007}. The CWS framework  enables distributed compression with random projections employing synchronous multiple access channels (MACs). With MAC,   individual nodes transmit their scaled observations to the fusion center coherently.  This architecture is depicted  in Fig. \ref{fig_MAC} for one MAC transmission.
With this model, the observation at the  $j$-th  sensor, denoted by $x_j$, is multiplied by a scalar factor, and transmitted to the fusion center via a MAC  channel. The received signal at the fusion center after the $i$-th MAC transmission is given by,
\begin{eqnarray}
 y_i =\sum_{j=1}^L \mathbf A_{i,j} x_k +  v_i\label{MAC_AWGN_i}
\end{eqnarray}
where $\mathbf A_{i,k}$ is the scalar factor, and $v_i$ is the receiver noise. After $M < L$ such MAC transmissions, the observation vector at the fusion center has the form of \eqref{smv_model_noise}
%\begin{eqnarray}
%\mathbf y = \mathbf A \mathbf x + \mathbf v\label{MAC_AWGN}
%\end{eqnarray}
where the $(i,j)$-th element of  $\mathbf A$ is given by $\mathbf A_{i,j}$ for $i=1,\cdots,M$ and $k=1,\cdots,L$, $\mathbf x= [x_1, \cdots, x_L]^T$ and $\mathbf v = [v_1, \cdots, v_M]^T$. With this model, the fusion center receives a compressed version of the sparse (or compressible) signal $\mathbf x$. In worth noting that, to avoid ambiguity in the paper, in this section, we denote the signal under compression as a length $L$ vector in contrast to a length $N$ vector as considered in Section \ref{sec_temporal}.

\begin{figure*}
\centering
\includegraphics[width=4.5in]{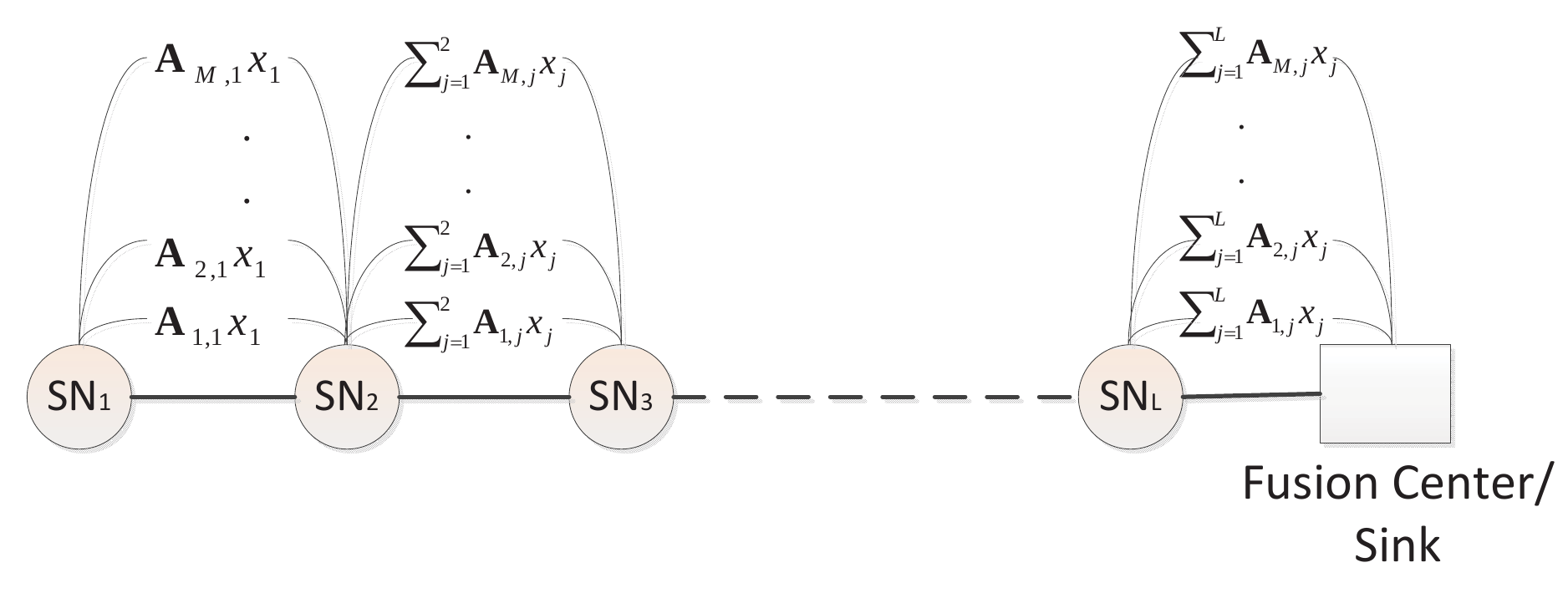}
 \caption{Compressive data gathering with multi-hop routing exploiting spatial sparsity \cite{Luo_TWC10,Luo_mobicom2009}}
\label{fig_DataGathering_Tree}
\end{figure*}

Note that the CWS architecture requires synchronization among multiple sensors during each MAC transmission. To alleviate this requirement, another  architecture widely considered is employing random projections  through multi-hop transmission \cite{Luo_TWC10,Luo_mobicom2009,Zheng_TWC13,Liu_PDS2015,Tang_TWC13}.   In this framework, the data aggregation process is implemented  with multi-hop routing as illustrated in Fig. \ref{fig_DataGathering_Tree}. In \cite{Luo_TWC10,Luo_mobicom2009,Tang_TWC13}, a tree based architecture is used to implement the multi-hop routing protocol so that the fusion center receives the observation vector $
\mathbf y = \mathbf A \mathbf x
$.
In the baseline multihop routing approach (without any compression), the $j$-th sensor transmits its reading $x_j$ along with the messages received from the previous node to the next node, thus, the nodes located closer to the  sink consume a large amount of  energy. In contrast, in the CS based multihop routing approach in Fig. \ref{fig_DataGathering_Tree}, each node has to transmit only $M = \mathcal O(k\log L)$ messages where $k$ is the sparsity index with respect to an appropriate basis as defined before.

In both single-hop or multi-hop architectures,   the problem of compressive data gathering exploiting spatial sparsity can be formulated as  a CS recovery  problem at the fusion center with a SMV. The works reported in \cite{haupt_SPM2008,Bajwa_IT2007} and  \cite{Luo_mobicom2009} are  the first few  papers that  demonstrated the  savings  in communication and computations costs using direct   application of   CS techniques in  large scale  networks in contrast to the  traditional  sample-then-process approach. Robustness of  CS based data gathering with this set-up in the presence  of link failures and outlying sensor readings  is further studied in \cite{Tang_TWC13}.   While  directly applying   random projections to compress spatial data  saves communication power costs compared to  gathering raw data, still the transmission power costs can be high since all the nodes participate in the data aggregation process. In particular, when a dense Gaussian matrix is used  for compression in the  multi-hop routing architecture, the communication cost in terms of the total
number of message transmissions and/or the maximum
number of message transmissions of any single node can be high.   In order to take limited communication and energy  resources into account in CS based data gathering exploiting spatial sparsity in large scale sensor networks, further research has been done mainly focusing on controlling the amount of information transmitted by each node (equivalently designing projection matrices) and developing reconstruction algorithms.

\subsubsection{Centralized processing: Use of sparse matrices for spatial data gathering}\label{sec_sparse}
\begin{figure*}
\centering
\includegraphics[width=4.0in]{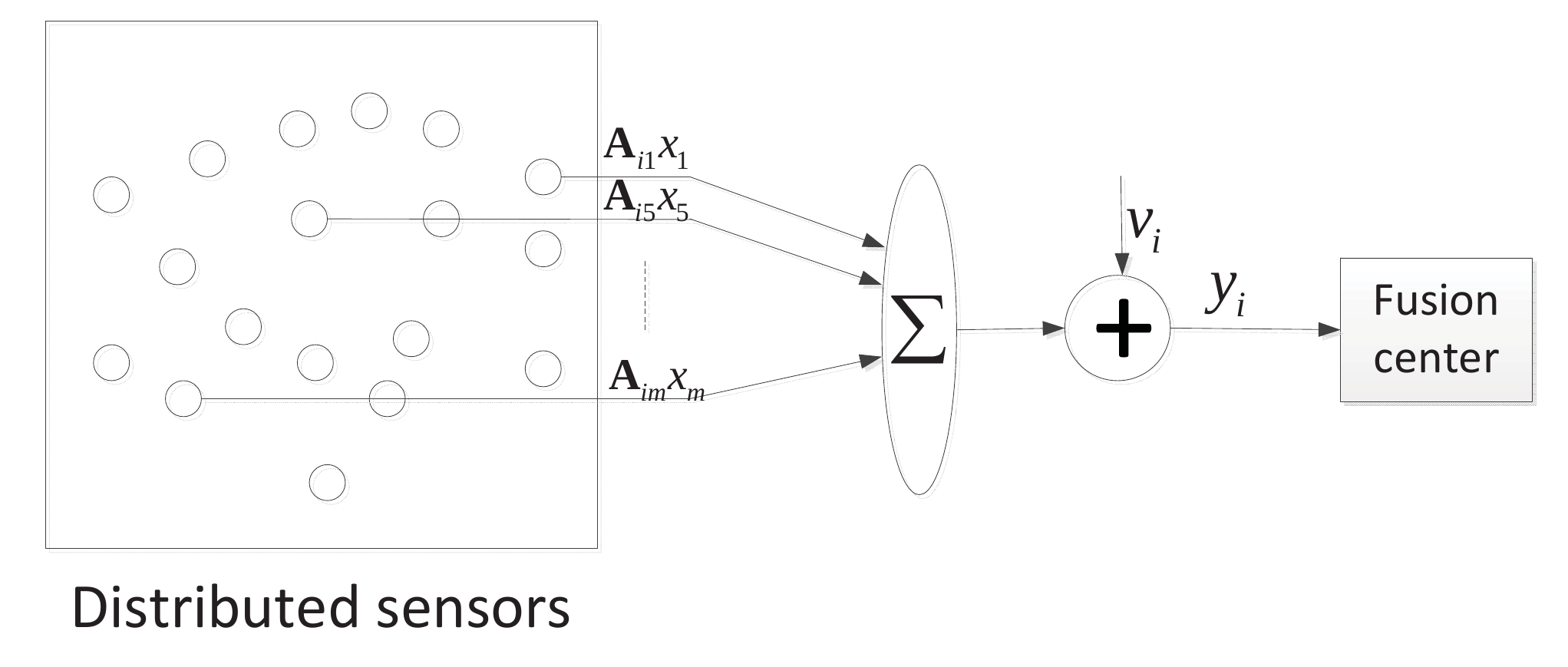}
 \caption{Compressive Wireless Sensing  with sparse random projections; only few sensors transmit during a given MAC transmission.}
\label{fig_MAC_sparse}
\end{figure*}

When dense random projections such as Gaussian random variables are used,  each node has to forward all of its scaled observations which can incur a significant transmission power cost. When the projection matrix is made sparse, not all the sensors need to transmit all the local data under both single-hop  and multi-hop architectures. The use of sparse random projections has the potential of reducing both communication and computational costs at sensor nodes. In CS theory, the signal recovery guarantees with sparse random matrices  are widely  discussed  \cite{wang_IPSN07,Shen_Sensor2013,Yang_TSP13,Wimalajeewa_TSIPN15,wang_TIT10,
Achlioptas_JCSC2003,Gilbert_Proc2010,Li_KDD2006}.    It is noted that many desirable properties that need to be satisfied to enable reliable reconstruction are violated when the projection matrix is made very sparse. Nevertheless, the authors in \cite{Li_KDD2006} have shown that very sparse random matrices can be used for data compression with a small recovery  performance loss.
Some widely considered sparse random projections are sparse Bernoulli and sparse Gaussian matrices as defined below. In sparse Bernoulli matrices, the $(i,k)$-th element of $\mathbf A$ is drawn from  \cite{wang_IPSN07}
\begin{eqnarray}
\mathbf A_{i,k} =\sqrt{\frac{1}{\gamma}} \left\{
\begin{array}{cccc}
1~ & \mathrm{with} ~ \mathrm{prob} & \frac{\gamma}{2}\\
0 ~& \mathrm{with} ~ \mathrm{prob} &1- \gamma\\
-1 ~& \mathrm{with} ~\mathrm{ prob} &\frac{\gamma}{2}
\end{array}\right.\label{sparse_Bern}
\end{eqnarray}
while in sparse Gaussian matrices $\mathbf A_{i,k}$ is chosen as   \cite{wang_TIT10}
\begin{eqnarray}
\mathbf A_{i,k} = \left\{
\begin{array}{cccc}
\mathcal N(0,\frac{1}{\gamma})~ & \mathrm{with} ~ \mathrm{prob} & \gamma\\
0 ~& \mathrm{with} ~ \mathrm{prob} &1- \gamma\label{sparse_Gauss}
\end{array}\right.
\end{eqnarray}
with $0<\gamma<1$. The CWS scheme with sparse random projections in a centralized architecture  is depicted in Fig. \ref{fig_MAC_sparse} for one MAC transmission.

Using the sparse  matrix \eqref{sparse_Bern}, it has been shown that it is sufficient to have $\mathcal O(poly(k,\log L))$ random projections (equivalently  MAC transmissions) to reconstruct the original signal with high probability  when only $\mathcal O(\log L)$ nodes on  an average transmit  their scaled observations during a given MAC transmission \cite{wang_IPSN07}.  Instead of randomly selecting participating nodes as in \cite{wang_IPSN07}, design of which nodes to transmit or equivalently constructing routing trees minimizing the
overall energy consumption has been  considered in \cite{Ebrahimi_COML2013}. While the optimal solution is NP-hard, several suboptimal approaches have been proposed with significantly less amount of data to be transmitted compared to the case where raw data forwarding. The minimum number of MAC transmissions required to enable signal reconstruction with high probability in the presence of fading has been analyzed by \cite{Yang_TSP13} using the  matrix \eqref{sparse_Bern} while the authors in \cite{Wimalajeewa_TSIPN15} have provided a similar analysis with the matrix \eqref{sparse_Gauss} assuming phase coherent transmission.  Impact of fading channels on signal reconstruction is further discussed in Section \ref{sec_fading}.

In \cite{Luo_TWC10}, design of the matrix $\mathbf A$ so that the total communication cost kept  under a desired level has been  investigated. The matrix should be designed so  that reliable  recovery is guaranteed. In particular, the authors  have proposed to  split the measurement matrix and incorporated sparsity to   one side so that the RIP property is not significantly  violated. Thus, the designed matrix with smaller communication overhead is able to provide closer performance to the case  using a dense matrix which consumes a large communication burden.  The authors in \cite{Zheng_TWC13} have also considered multi-hop routing based CS data gathering,  considering more general network models. In particular, a single sink as well as multi-sink models were considered. The capacity; the maximum rate that the sink can receive  the data generated at nodes, and the delay; the time between the data sampling at nodes and the receiving at the sink were analyzed with both single sink and multi-sink models  exploiting  sparse  random projections.  In \cite{Wu_TWC14},  the  measurement matrix has been made as sparsest as possible where each row contains only one element. Since this matrix is not capable of providing reliable  recovery guarantees when representing sensor data in a common transform basis, the authors have  designed the transform basis so that the corresponding projection matrix is capable of providing reliable recovery guarantees. While this method reduces the communication cost significantly it adds computational cost in designing the transform basis as required for signal reconstruction.  In \cite{Liu_PDS2015}, a  non-uniform version of the  sparse Bernoulli matrix \eqref{sparse_Bern}  has been  used where each column is generated using sparsity  parameter $0<\gamma_j <1$ for $j=1,\cdots,L$. Recovery guarantees  with such non-uniform sparse matrices have been established in \cite{Liu_PDS2015}.  In terms of recovery performance,  it is comparable to the case with  direct application of CS as in \cite{Luo_mobicom2009} with large $M$, however, it leads to a  performance degradation compared to \cite{Luo_mobicom2009} when $M$ is small. Incorporating practical constraints such as interference in adjacent transmissions into sparse projections based data gathering framework has been addressed in \cite{Ebrahimi_TWC16,Zheng_Access17}.

{Hybrid and adaptive approaches to design the measurement matrices  and/or the reconstruction algorithms   to minimize communication cost in data gathering exploiting spatial sparsity  are also attractive. Recent developments of adaptive CS are exploited in  these approaches which will be discussed next. }

\subsubsection{Centralized processing:  Hybrid and adaptive approaches for CS based spatial data gathering}\label{sec_adaptive}
In \cite{Xiang_SECON2011}, a hybrid data aggregation approach has been  proposed combining the energy efficiency
aspect of applying CS  to data collection
in WSNs focusing on multihop networking scenarios. In particular, by joint routing and compressed aggregation, the energy consumption  is reduced compared to applying CS directly as in \cite{Luo_mobicom2009}. Ideas of adaptive compressive sensing \cite{Indyk_FOCS2011}, where the goal is to design projections in an   iterative manner to maximize the amount of information and minimize the number of projections, have been  exploited in \cite{Chou_LCN2009}, to design projection matrices. In particular, the subsequent projections have been  designed  taking into both energy constraints and the information gain into account so that the  desired recovery performance is achieved at the sink. Simulated as well as testbed (to measure temperature) setups have been  used to   adaptively collect sensor data in \cite{Chou_LCN2009}. Adaptive techniques for CS based data gathering have also been explored in \cite{Wang_INFOCOM12,Aderohunmu_Sensor2015,Yin_Sensor2016}. In \cite{Wang_INFOCOM12}, both measurement scheme and the reconstruction approach are adaptively updated  to cope with the variation of sensed data as well as the application requirements.  Testbed data has been used to illustrate the variation of senor readings due to external event  and internal errors as well as to show the performance of the  proposed adaptive techniques. In \cite{Aderohunmu_Sensor2015},  a testbed for wildfire monitoring has been used to illustrate the performance of the proposed adaptive data compression scheme exploiting spatial sparsity. In \cite{Yin_Sensor2016}, a feed-back controlled adaptive technique has been proposed to cope with the variation  of sensor readings. The performance  has been evaluated using a testbed designed to monitor the  luminosity of the environment.

\begin{table*}
\renewcommand{\arraystretch}{1.3}
 %if using array.sty, it might be a good idea to tweak the value of
% \extrarowheight as needed to properly center the text within the cells
\caption{Centralized solutions  for  sparse signal recovery  exploiting  spatial  sparsity}
\label{table_spatial_sparse}
\centering
% Some packages, such as MDW tools, offer better commands for making tables
% than the plain LaTeX2e tabular which is used here.
\begin{tabular}{|l|l|l|}
\hline
Architecture & Features & References \\
\hline
\hline
Single-hop  & Use of MAC,  consider  AWGN channels   & \cite{haupt_SPM2008,Bajwa_IT2007}\\
$~$ & Use of MAC, consider fading channels & \cite{Yang_TSP13,Wimalajeewa_TSIPN15}\\
\hline
Multi-hop  & Direct application of  CS  & \cite{Luo_mobicom2009,Tang_TWC13}\\
$~$ & Design of sparse random matrices to meet communication constraints   & \cite{Li_KDD2006,Luo_TWC10,Zheng_TWC13,Wu_TWC14,Liu_PDS2015,Ebrahimi_TWC16,Zheng_Access17}\\
$~$ & Use adaptive, and hybrid approaches  & \cite{Chou_LCN2009,Wang_INFOCOM12,Aderohunmu_Sensor2015,Xiang_SECON2011,Yin_Sensor2016}\\
\hline
\end{tabular}
%\begin{enumerate}
%\item
%  \end{enumerate}
\end{table*}

In Table \ref{table_spatial_sparse}, we summarize the centralized data gathering approaches discussed above exploiting spatial sparsity.

\subsubsection{Decentralized solutions for  sparse signal recovery exploiting spatial sparsity}\label{sec_spatial_decentralized}
In contrast to forwarding the compressed data to a sink node for reconstruction, decentralized  implementation  is to estimate the  sparse vector of interest in a distributed/decentralized manner  when  a given node has  access to only local or  some partial information about the information that would go to a sink node in a centralized setting.  This requires the communication to be only within one-hop neighborhood of each node thus reducing the communication cost significantly compared to multi-hop transmission and long-range single-hop transmission as considered in Section \ref{sec_spatial}.

\begin{figure}[h!]
\centering
\includegraphics[width=2.0in]{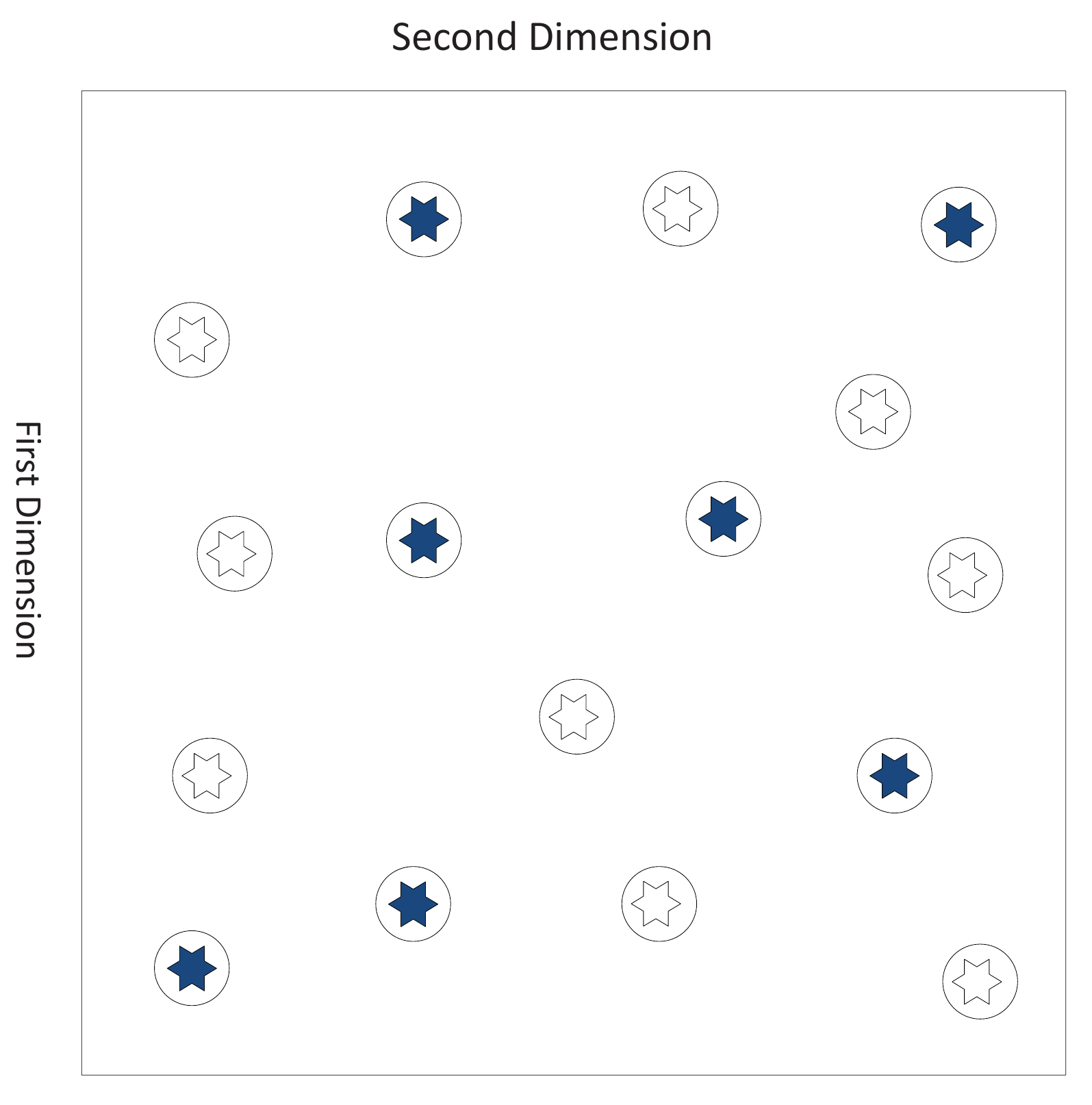}
 \caption{Sparse event monitoring;  (\cite{Yuan_PDS2015,Ling_TSP10}), locations of sources (stars) coincide with the locations of sensors (circles),  solid stars - active sources, void stars-inactive sources}
\label{fig_Sparse_event1}
\end{figure}

Most of the decentralized algorithms developed exploiting spatial sparsity  are application specific since the constraints,  available local information and the communication overhead  depend on the application scenario.
The work in \cite{Ling_TSP10} models the sparse event monitoring task as a sparse signal reconstruction   problem  where compression is achieved letting only few active sensors collect data.   With  $L$ distributed sensors  each having a position  denoted by $r_i$ for $i=1,\cdots,L$, the sources of events are  confined  to
sensor points; i.e., an  event occurs only at a sensor
point as shown in Fig. \ref{fig_Sparse_event1}.
The magnitude of the event at the $r_i$-th location  is denoted by a positive scalar $s_i$. If  no event occurred at the location $r_i$, $s_i=0$,  then those sensors are called inactive sensors. Forming a vector $\mathbf s= [s_1, \cdots, s_L]^T$, the measurement at the sensor located at $r_i$ is given by,
\begin{eqnarray}
y_i = \underset{j}{\sum} \boldsymbol\Psi_{i,j} s_j + v_i
\end{eqnarray}
where $ \boldsymbol\Psi_{i,j}$ denotes the influence of a unit-magnitude event at the position $r_j$ on the sensor position at $r_i$ for $i,j=1,\cdots, L$. The observation vector at $M < L$ active sensors in  vector-matrix notation has the form of \eqref{smv_model_noise}
%\begin{eqnarray}
%\mathbf y = \mathbf A \boldsymbol\Psi \mathbf s + \mathbf v\label{Sparse_event_detection_M1}
%\end{eqnarray}
where $\mathbf y =[y_1, \cdots, y_M]^T$, $\mathbf A$ in \eqref{smv_model_noise} is replaced by $\mathbf A \boldsymbol\Psi$ with  $\boldsymbol\Psi$ being  an ${L\times L}$ matrix where $\boldsymbol\Psi_{i,j}$ is  the $(i,j)$-th element, and  $\mathbf A$ being  an $M\times L $ matrix which selects the $M $ rows of $\boldsymbol\Psi$ corresponding to active sensors and $\mathbf v = [v_1, \cdots, v_M]^T$.
In this formulation, the problem of estimating  the active sources  is formulated as estimating the sparse vector $\mathbf s$ so that \cite{Ling_TSP10}
 \begin{eqnarray}
 \underset{\mathbf s}{\arg\min} \frac{\lambda}{2}||\mathbf A \boldsymbol\Psi \mathbf s - \mathbf y||_2^2  + ||\mathbf s ||_1 ~\mathrm{such} ~\mathrm{that} ~\mathbf s \geq \mathbf 0\label{l1_reg_nonnegative}
 \end{eqnarray}
 where $\lambda$ is a penalty parameter which is a generalization of the LASSO formulation. The  main idea of the decentralized implementation in \eqref{l1_reg_nonnegative} is to estimate the $s_i$ at the $i$-th node and the elements corresponding to the inactive sensors in the $i$-th node's one hop neighborhood by collaborating with the neighboring active nodes exploiting consensus optimization and ADM ideas. In this framework, in contrast to estimating the whole length $N$ sparse vector at each node, the coefficients corresponding to itself and neighboring nodes are estimated. This reduces the amount of information to be exchanged among nodes in the  one hop neighborhood.  In \cite{Yuan_PDS2015}, a similar  problem formulation as in \cite{Ling_TSP10} has been  considered  where the authors have proposed to use partial consensus    based and the Jacobi approaches  to further reduce the communication cost in the decentralized implementation.

\subsection{Data Gathering Exploiting Spatio-Temporal Sparsity}\label{sec_spatio_temporal}
In Section \ref{sec_temporal}, we considered the case were random projections are used to compress a single high dimensional sparse vector  of time samples collected at a given node. The goal was to simultaneously estimate  multiple such  high dimensional sparse signals exploiting joint structures. In particular, the sparsity of the signal is considered with respect to the transform basis for temporal data.  On the other hand, in Section \ref{sec_spatial}, the spatial samples collected at multiple nodes at a given time instant were compressed  using random projections in a distributed manner. In this case, sparsifying basis was taken with respect to the spatial data.   In both Sections \ref{sec_temporal} and \ref{sec_spatial}, sparsifying basis was considered  with respect to a single vector.  In many applications, network data generally exhibits spatio-temporal correlations \cite{Ali_SENSORCOMM2009}, thus, data compression exploiting sparsity in both dimensions (spatial and temporal) leads to better performance than the case where compression is done considering only one dimension \cite{Cheng_TWC13}. {In the following, we discuss CS techniques that can be utilized to process high dimensional data exploiting spatio-temporal sparsity. }

Recall that $\mathbf X$ defined in Section \ref{sec_temporal} is a $N\times L$ matrix in which each column contains the length-$N$ data vector (uncompressed)  collected at a given node. When exploiting spatio-temporal sparsity, it is desired that $\mathbf X$ is compressed both row- and column-wise. When implementing such compression schemes in WSN settings, there have been research efforts which exploit some concepts developed in CS theory for sparse matrix compression \cite{Cheng_Globecom2010,Cheng_TWC13,Quan_COML2016,Leinonen_TWC15}

\subsubsection{Use of  matrix completion techniques}\label{sec_matrix_completion}
In \cite{Cheng_Globecom2010,Cheng_TWC13}, the authors have exploited the ideas of matrix completion \cite{Candes_FCM2009} inspired by  CS theory to develop efficient data gathering schemes exploiting spatio-temporal sparsity. In particular, the matrix $\mathbf X$ can be assumed to be low rank \cite{Vuran_Else2004} in the presence  of spatio-temporal correlations which can be recovered reliably  from a compressed version   of it  (or a subset of its entries)  using  matrix completion ideas \cite{Candes_FCM2009}.   Low rank feature of data  collected by WSNs has been demonstrated using   testbed experiments in  \cite{Cheng_TWC13}. Mathematically the $\mathbf X$ can be  found by  solving the optimization problem
\begin{eqnarray}
\underset{\mathbf X}{\min} ~\mathrm{rank} (\mathbf X)~ \mathrm{such} ~\mathrm{that }~f_A(\mathbf X) = \mathbf Y \label{rank_min}
\end{eqnarray}
where $f_A(\cdot)$ denotes a linear operation  and $\mathbf Y$ is the received data matrix at the sink.  Solving \eqref{rank_min} in its current form  is NP hard, thus there have been several attempts to find approximate solutions for \eqref{rank_min} where nuclear norm minimization as discussed in \cite{Recht_SIAM2010} is shown to be a promising approach. In the data gathering scheme presented in  \cite{Cheng_TWC13}, each sensor forwards  its observed data to the sink with a certain probability leading to sparsity in the matrix $\mathbf X$.  By defining  $f_A(\cdot) = \mathbf X\odot \mathbf Q$ where $\mathbf Q$ is a $N\times L$ matrix consisting of $1$'s and $0$'s, the data gathering problem is solved via the  nuclear norm minimization approach which has been  shown to  outperform the  case where only spatial sparsity is exploited as  in the plain CS setting \cite{Luo_mobicom2009}. The performance evaluation is done using testbed data reported in \cite{Test_bedINTEL} as well as with a testbed setup implemented by the authors in   \cite{Cheng_TWC13} to collect temperature, light, humidity, and
voltage data.

\subsubsection{Use of structured/Kronecker CS}
Another approach considered to exploit spatio-temporal sparsity in data gathering is to exploit ideas in Kronecker CS \cite{Duarte_TIP2012}. The Kronecker CS framework can be used to  combine the individual sparsifying bases in both spatial and temporal domains to a  single transformation matrix. In particular, if each column in $\mathbf X$ has a sparse representation in the basis $\boldsymbol\Phi_c\in \mathbb R^{N\times N}$ and each row  has a sparse representation in the basis $\boldsymbol\Phi_r\in \mathbb R^{L\times L}$, the vectorized \footnote{vectorized version is obtained by stacking columns of $\mathbf X$ one after the other.} version of $\mathbf X$ denoted by $\mathrm{vec}(\mathbf X)$ has a sparse representation in  $\boldsymbol\Phi_r \otimes \boldsymbol\Phi_c \in \mathbb R^{NT\times NT}$ \cite{Duarte_TIP2012}.   Kronecker CS  ideas have been used for spatio-temporal data gathering in \cite{Gong_COML2015,Quan_COML2016,Leinonen_TWC15,Li_WCL18} with enhanced performance compared to the use of CS only considering  a single domain. It  is noted that with the Kronecker based formulation, the communication and computational complexities  of the data transmission and  recovery algorithms increase  due  to the need of handling a $NT\times NT$  matrix. In order to reduce communication  complexity, a Kronecker based approach has been proposed in \cite{Quan_COML2016} where nodes communicate with only limited number of nodes in data forwarding. In  \cite{Leinonen_TWC15}, a sequential approach has been discussed  which exploits Kronecker sparsifying bases (in spatial and temporal domains) with improved performance compared to using Kronecker CS ideas \cite{Duarte_TIP2012} directly.

\begin{table*}
\renewcommand{\arraystretch}{1.3}
 %if using array.sty, it might be a good idea to tweak the value of
% \extrarowheight as needed to properly center the text within the cells
\caption{Data gathering exploiting spatio-temporal sparsity}
\label{table_spatio-temporal}
\centering
% Some packages, such as MDW tools, offer better commands for making tables
% than the plain LaTeX2e tabular which is used here.
\begin{tabular}{|l|l|l|}
\hline
Technique & Features & References \\
\hline
\hline
Matrix completion techniques  & Centralized processing, exploit low-rank property     & \cite{Cheng_Globecom2010,Cheng_TWC13}\\
$~$ & of spatio-temporal data, nuclear norm minimization & $~$\\
\hline
structured/Kronecker CS  &Centralized processing, combine spatial and temporal       & \cite{Gong_COML2015,Quan_COML2016,Leinonen_TWC15,Li_WCL18}\\
$~$ & sparsifying bases to a  single transformation matrix & $~$\\
\hline
\end{tabular}
%\begin{enumerate}
%\item
%  \end{enumerate}
\end{table*}
{Advances of CS techniques  discussed above  to solve the data gathering problem exploiting spatio-temporal sparsity are  summarized in Table \ref{table_spatio-temporal}. }

{Irrespective of the form of sparsity exploited, the data gathering problem as discussed in this section deals with complete signal reconstruction be it  centralized or distributed.  In the next section, we discuss other applications of WSNs that do not necessarily require signal reconstruction, however,  CS can still  be utilized }

\section{CS for  Distributed Inference  }\label{sec_inference}
In WSNs, CS techniques can be used as a means of  data compression while the end result is not signal reconstruction but  solving  an inference problem. For example, in order to solve  a variety of inference problems such as detection, classification, estimation of parameters  and tracking, it is sufficient to construct  a reliable decision statistic  based on compressed data  without completely recovering the original  signal. Beyond the standard CS framework, this requires the investigation of different metrics for performance analysis and quantification of the amount of information
preserved under compression to obtain a reliable inference decision. Further, in contrast to
complete signal reconstruction, sparsity prior is not necessary and  the performance and the specific design principles depend on how
the signals or the noise  parameters are modeled. Thus, research on  the applications  of  CS to solve such inference problems under resource constraints is attractive in WSNs. In the following, we discuss detection, classification, and and parameter estimation problems with CS.

\subsection{Compressive Detection}
Detection is one of  the fundamental  tasks  performed by WSNs \cite{Varshney_B1}. In order to solve a detection problem efficiently, a decision statistic needs to be computed by intelligently combining multisensor  data under resource constraints.  In order to minimize the amount of information to be transmitted by sensor nodes, processing  sensor data locally is of great importance. The overall decision statistic further depends on how the signals of interest and noise are modeled.   In the following, we discuss how CS can be beneficial in solving  detection problems. {Based on the specific application and the approach employed, we classify the existing CS based detection techniques into five  categories which will be presented in Subsections \ref{sec_det_sparse_event} to \ref{sec_det_matrix_design}.}
%In particular, CS ideas can be exploited in  reducing communication overhead as well as in constructing decision statistics exploiting sparsity in different dimensions.

\subsubsection{Sparse events detection exploiting spatial sparsity}\label{sec_det_sparse_event}
 An immediate application of CS in detection is to exploit spatial sparsity in sparse  event detection problems.
Detection of the presence of rare events by a sensor networks has applications in diverse areas such as   environment monitoring  and alarm systems \cite{Wittenburg_MCOM12}. When the number of active events is  lees than the number of all possible events, the event detection problem can be reformulated as a sparse recovery problem.   CS ideas have been exploited in sparse event detection in \cite{Yuan_PDS2015,Ling_TSP10,meng_CISS09,Alwakeel_wcnc2014}. Spatial sparsity can be exploited in sparse event  detection in several ways.

In \cite{meng_CISS09},   the sparse event detection problem is formulated  in the following sense. Let there be altogether $K$ sources in which $k$ (out of $K$) are active. In particular, the $k$ events are assumed to occur  simultaneously.
 \begin{figure}[h!]
\centering
\includegraphics[width=2.0in]{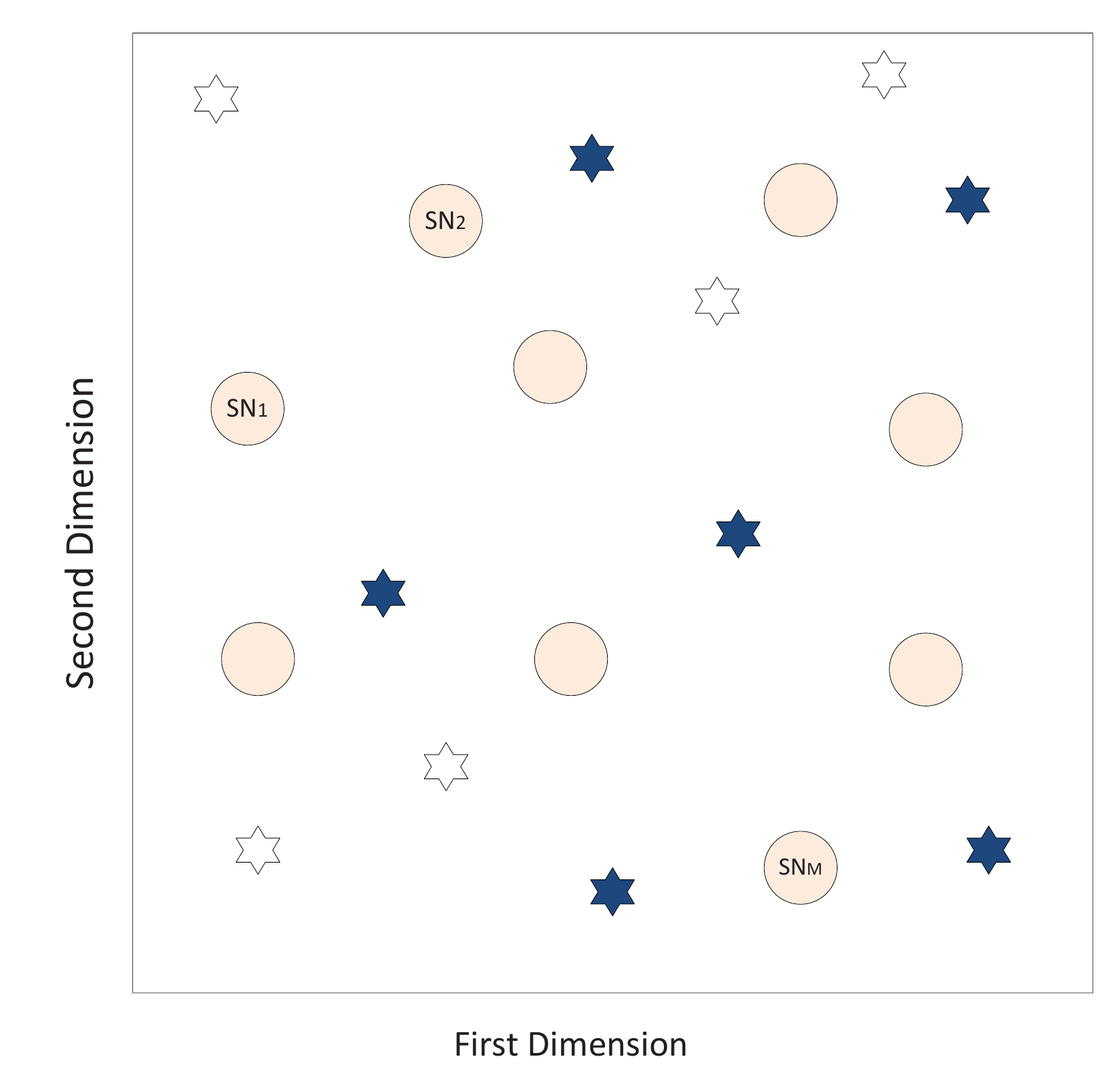}
 \caption{Sparse event monitoring; (\cite{meng_CISS09}), dark stars-active sources, void  stars-inactive sources, circles-sensor nodes}
\label{fig_Sparse_event2}
\end{figure}
The measurement vector at $M$ active  sensors (out of $L$) has the form of \eqref{smv_model_noise}
%\begin{eqnarray}
%\mathbf y = \mathbf A \mathbf x + \mathbf v\label{obs_sparseEvent_M2}
%\end{eqnarray}
where $\mathbf x$ is a sparse vector with $\mathbf x[i]\in \{0,1\}$. The $(i,j)$-th element  of  $\mathbf A$ is given by
\begin{eqnarray}
\mathbf A_{i,j} = r_{i,j}^{-\alpha/2}|h_{i,j}|\label{A_sparse_EventM2}
 \end{eqnarray}
 where  $r_{i,j}$ denotes the distance from the $i$-th sensor to the $j$-th source, $\alpha$ is the propagation loss factor and $h_{i,j}$ is the fading coefficient of the channel between the $i$-th sensor and  the $j$-th source.  In this framework, the sparse event monitoring problem reduces to estimating  the sparse vector $\mathbf x$ from \eqref{smv_model_noise} when the elements of $\mathbf A$ as  given in \eqref{A_sparse_EventM2}.  Note that here $\mathbf A$ is not a user defined random matrix satisfying RIP properties as desired by the standard CS framework. In particular, the randomness arises due to the random locations and fading coefficients.   The authors in \cite{meng_CISS09},  have  developed  Bayesian algorithms  exploiting marginal likelihood maximization for sparse event detection. In \cite{Alwakeel_wcnc2014}, the problem of sparse event detection has been  addressed in an adaptive manner using sequential compressive sensing (SCS). SCS based event detection is shown to outperform the Bayesian approach considered in \cite{meng_CISS09} in the low SNR region. The sparse event detection problem  has been formulated  from the perspective of coding
theory  by modeling the detection problem as a
decoding of Analog Fountain Codes (AFCs) in \cite{Shirva_Globecom2014}.
 The decentralized approach proposed in  \cite{Ling_TSP10} for signal recovery exploiting spatial sparsity  as discussed in Section \ref{sec_spatial_decentralized} can also be used to event detection when  the events are assumed to be confined to sensor locations.

\subsubsection{Detection exploiting temporal sparsity}
Consider the following  detection problem with the time samples collected at the $j$-th node given by
\begin{eqnarray}
\mathcal H_1: ~ \mathbf x_j &=& \boldsymbol\theta_j + \mathbf v_j \nonumber\\
\mathcal H_0: ~ \mathbf x_j &=& \mathbf v_j\label{obs_uncomp}
\end{eqnarray}
where $\boldsymbol\theta_j\in \mathbb R^N $ is the (unknown) signal   observed  by the $j$-th node and $\mathbf v_j$ is the additive noise for $j=1, \cdots, L$. The hypotheses $\mathcal H_1$ denotes that the signal is present while $\mathcal H_0$ denotes the signal is absent.  Let the signal to be detected be sparse in the  basis $\boldsymbol\Psi$ so that $\boldsymbol{\theta}_j = \boldsymbol\Psi \mathbf s_j$ with $\mathbf s_j$ having only a few nonzero elements. With the common support set model, the coefficients $\mathbf s_j$ for $j=1,\cdots, L$  share  the same support.
 Instead of transmitting $\mathbf x_j$'s, let the nodes transmit only a compressed version by applying random projections to the observations. When the detection problem is solved using compressed data, it is important to understand how to compute decision statistics along with their performance.  After compression,   the detection problem needs to be solved based on
\begin{eqnarray}
 \mathbf y_j &=& \mathbf A_j\mathbf x_j \label{obs_comp}
\end{eqnarray}
for $j=1,\cdots,L$.
In particular, the goal is to decide between hypotheses   $\mathcal H_1$ and  $\mathcal H_0$ based on (\ref{obs_comp}).
The detection problem in \eqref{obs_comp} can be expressed as:
\begin{eqnarray}
\mathcal H_1: ~ \mathbf y_j &=&   \mathbf B_j \mathbf s_j + \tilde{ \mathbf v}_j \nonumber\\
\mathcal H_0: ~ \mathbf y_j &=& \tilde{ \mathbf v}_j \label{det_comp}
\end{eqnarray}
for $j=1,\cdots, L$ where  $\tilde{ \mathbf v}_j=\mathbf A_j \mathbf v_j\sim\mathcal (\mathbf 0, \sigma_v^2 \mathbf I_M)$ when $\mathbf A\mathbf A^T = \mathbf I_M$ and $\mathbf v_j\sim \mathcal \sigma_v^2 \mathbf I_N$.
When the support of $\mathbf s_j$'s denoted by  $\mathcal U$ is exactly known,  (\ref{det_comp}) reduces to
\begin{eqnarray}
\mathcal H_1: ~ \mathbf y_j &=& \mathbf B_j(\mathcal U)\mathbf s_j(\mathcal U) +\tilde{ \mathbf v}_j\nonumber\\
\mathcal H_0: ~ \mathbf y_j &=&\tilde{ \mathbf v}_j \label{obs_support_known}
\end{eqnarray}
for $j=1,\cdots,L$ where $ \mathbf B_j(\mathcal U)$ denotes the $M\times k$ submatrix of $\mathbf B_j=\mathbf A_j \boldsymbol\Psi$ in which columns are indexed by the ones in $\mathcal U$, and $\mathbf s_j(\mathcal U)$ is a $k\times 1$ vector containing nonzero elements in $\mathbf s_j$ indexed by $\mathcal U$ for $j=1,\cdots,L$.  When  $ \mathbf B_j(\mathcal U)$  is known, (\ref{obs_support_known}) is the subspace detection problem which has been addressed previously \cite{Scharf_TSP94,Scharf_ASAP03,jin_TSP2005}. Depending on how the unknown coefficient vector $\mathbf s_j(\mathcal U)$ is modeled, different detectors have been  proposed. In  \cite{Scharf_TSP94}, a generalized likelihood ratio test (GLRT) based detector has been  proposed when $\mathbf s_j(\mathcal U)$ is assumed to be deterministic. In \cite{Scharf_ASAP03}, the analysis has been  extended to the case  when $\mathbf s_j(\mathcal U)$ is modeled as  random. The problem with multiple observation vectors  has been addressed in \cite{jin_TSP2005} where the authors have proposed adaptive subspace detectors   when the coefficients  $\{\mathbf s_j(\mathcal U)\}_{j=1}^L $ follow first and second order Gaussian models.

In the case of sparse signal detection, it is unlikely that the exact knowledge of $\mathcal U$ is available \emph{a priori}.  In other words,  sparse signal detection needs to be performed when  $\mathcal U$ is unknown. When the sparsity prior is ignored, one of the common approaches would be to consider the GLRT where  ML estimator of $\mathbf s_j$ is found as  \cite{kay_det}
 \begin{eqnarray}
 \hat {\mathbf s}_j = (\mathbf B_j^T \mathbf B_j)^{-1} \mathbf B_j^T \mathbf y_j.\label{ML_estimate}
 \end{eqnarray}
 When $\mathbf s_j$ is sparse, the ML estimate is inaccurate since it is not capable of providing a sparse solution. Motivated by CS theory,  some sparsity aware algorithms have been developed to detect sparse signals based on  (\ref{det_comp}) by finding a better estimate for $ \hat {\mathbf s}_j $ for $j=1,\cdots,L$ \cite{haupt_ICASSP07,duarte_ICASSP06,Wimalajeewa_ICASSP13,Gang_globalsip14,Wimalajeewa_tsipn16} with enhanced performance compared to the ML based approach.
  In particular,  the standard OMP algorithm has been  modified in \cite{duarte_ICASSP06} to detect the presence of a sparse signal based on a SMV.  In \cite{Wimalajeewa_tsipn16}, the sparse signal detection problem has been  addressed with MMV. The authors have derived the minimum fraction of the support set to be estimated to achieve a desired detection performance. Further, distributed algorithms for detection  with  only partial support set estimation via OMP are  developed.   In \cite{Wimalajeewa_ICASSP13}, a heuristic algorithm has been  proposed for sparse signal detection in a decentralized manner based on partial support set estimation via OMP  at individual nodes.   Sparse detection problem in a Bayesian framework has been treated in \cite{Hariri_SP2017} without complete signal reconstruction.

\begin{table*}
\renewcommand{\arraystretch}{1.3}
 %if using array.sty, it might be a good idea to tweak the value of
% \extrarowheight as needed to properly center the text within the cells
\caption{CS based solutions for detection in WSNs}
\label{tab_detection}
\centering
% Some packages, such as MDW tools, offer better commands for making tables
% than the plain LaTeX2e tabular which is used here.
\begin{tabular}{|l|l|l|}
\hline
Problem & Features & References \\
\hline
\hline
Sparse event detection  & Formulate the sparse event detection problem as a sparse signal   & \cite{Yuan_PDS2015,Ling_TSP10,meng_CISS09,Alwakeel_wcnc2014,Shirva_Globecom2014}\\
$~$ &  reconstruction problem, exploit spatial sparsity & \\
\hline
Detection of temporal sparse signals & Compute decision statistics based on complete/partial   & \cite{haupt_ICASSP07,duarte_ICASSP06,Wimalajeewa_ICASSP13}\\
$~$ & signal reconstruction, compression at each node, exploit temporal sparsity  & \cite{Gang_globalsip14,Wimalajeewa_tsipn16}\\
\hline
Detection of non-sparse signals & Compression at each node,  no signal reconstruction  & \cite{davenport_JSTSP10,Kailkhura_TSP16,Wimalajeewa_TSP17,Zheng_icc11}\\
\hline
Detection of subspace signals & Compression at each node, known subspaces as well as   & \cite{Rao_icassp2012,Razavi_TSP2016,Wang_ICASSP08}\\
$~$ & unknown subspaces are considered & $~$\\
\hline
Design of measurement matrices & Optimize for measurement matrices
to achieve a given detection performance   &  \cite{Zahedi_Phys12,Bai_WCS2013,Kailkhura_WCL16} \\
\hline
\end{tabular}
%\begin{enumerate}
%\item
%  \end{enumerate}
\end{table*}

\subsubsection{Detection of non-sparse  signals}
Let us revisit the detection problem presented in \eqref{obs_uncomp}. When the signals to be detected,  $\boldsymbol\theta_j$'s are known, the optimal detector which minimizes the average probability of error is given by the matched filter. Consider the same problem in the compressed domain based on \eqref{obs_comp}. This problem has been treated with a single sensor in \cite{davenport_JSTSP10}.  When $\boldsymbol\theta_j$'s are  known and assuming $\mathbf v_j \sim \mathcal N(0, \sigma_v^2 \mathbf I)$, the decision statistic of the matched filter, $\Lambda_c$, in the  compressed domain is given by
\begin{eqnarray}
\Lambda_c = \frac{1}{\sigma_v^2}\sum_{j=1}^L\mathbf y_j^T (\mathbf A_j \mathbf A_j^T)^{-1} \mathbf A_j\boldsymbol\theta_j.
\end{eqnarray}
This results in the following probability of detection of the NP detector with probability of false alarm less than $\alpha_0$ \cite{Kailkhura_TSP16}:
\begin{eqnarray}
P_d^c  = Q\left(Q^{-1}(\alpha_0) - \frac{1}{\sigma_v}\sqrt{\sum_{j=1}^L ||\mathbf P_{\mathbf A_j^T} \boldsymbol\theta_j||_2^2}\right) \label{PD_comp}
\end{eqnarray}
where $\mathbf P_{\mathbf A_j^T}  = \mathbf A_j^T (\mathbf A_j \mathbf A_j^T)^{-1} \mathbf A_j$ and $Q(\cdot)$ denotes the Gaussian Q-function. When $\mathbf A_j$ is selected to be random and an orthoprojector so that $\mathbf A_j \mathbf A_j^T = \mathbf I$, \eqref{PD_comp} can be approximated by
\begin{eqnarray}
P_d^c  \approx  Q\left(Q^{-1}(\alpha_0) - \sqrt{\frac{M}{N} \mathrm{SNR}} \right)
\end{eqnarray}
where $\mathrm{SNR} = \frac{\sum_{j=1}^L||\boldsymbol\theta_j||_2^2}{\sigma_v^2}$.
With uncompressed observations, the matched filter results in the following probability of detection,  $P_d^u$:
\begin{eqnarray}
P_d^u = Q\left(Q^{-1}(\alpha_0) - \sqrt{\mathrm{SNR}} \right).
\end{eqnarray}
Thus,  the impact of performing  known  signal detection in the compressed domain appears on  the probability of detection via the argument of the $Q$ function. As discussed in \cite{davenport_JSTSP10,Kailkhura_TSP16},  when SNR is large, compressive detection is capable of providing similar performance as that of  the uncompressed detector.  In particular, transmitting only a compressed version of observations by each node does not, under certain conditions,  result in performance loss  compared to transmitting raw data. Thus, when the problem is detection (but not exact signal recovery), the CS measurement scheme  can still be  beneficial even without having the sparsity prior.  In \cite{Kailkhura_TSP16}, the performance of the compressed detector  when $\boldsymbol\theta_j$'s are random  has also been discussed. In \cite{Cao_Info2014}, the performance analysis of detection in a Bayesian framework with  unequal probabilities for hypotheses has been  presented. In  recent papers  \cite{Wimalajeewa_TSP17,Wimalajeewa_CAMSAP2017}, the authors have considered the problem of detection with multimodal dependent data analyzing  the potential of CS in capturing the second order statistics of uncompressed data to compute decision statistics for detection. Performance analysis of sequential detection in the compressed domain has been considered in \cite{Zheng_icc11} with a single as well as multiple sensors.

\subsubsection{Detection of subspace  signals}\label{sec_det_subspace}
Compressing time samples via random projections at each node to detect a variety of signals lying in a known low dimensional subspace (without complete  signal reconstruction)   has been investigated by several other authors.  In \cite{Rao_icassp2012}, the detection performance of    random sparse signals has been  derived assuming that the subspace  in which the signal is sparse is known.
Subspace signal detection with compress data  under realistic scenarios has been treated in \cite{Razavi_TSP2016}. Decision statistics have been derived when in teh presence of  unknown noise variance and imprecise measurements.    The results have been extended to  the case when the signal of interest lies in a union of subspaces.  The authors in  \cite{Wang_ICASSP08} have considered  the sparse signal detection problem assuming the signal to be detected  lies in a known subspace. When the corresponding subspace is unknown, detection is performed after estimating the subspace using some training data.

\subsubsection{Design of measurement matrices for detection}\label{sec_det_matrix_design}
When exploiting CS in signal compression to solve detection problems, random matrices that satisfy RIP properties are widely employed.   However, when the problem is to perform  detection, it is interesting to see if the same conditions as required for complete signal reconstruction are necessary for the measurement matrices.
Design of measurement matrices for compressive detection so that a given objective function is optimized  has been  considered in \cite{Zahedi_Phys12,Bai_WCS2013,Kailkhura_WCL16}. In \cite{Zahedi_Phys12}, measurement matrices are designed so that  the worst case SNR and the average minimum SNR are maximized. In \cite{Kailkhura_WCL16}, the authors have considered the probability of detection of the NP detector as the objective function and shown  that the optimal measurement matrices depend on the signal being detected. Detection with the designed measurement matrices  in  \cite{Zahedi_Phys12,Bai_WCS2013,Kailkhura_WCL16} is  shown to outperform that with random measurement matrices satisfying RIP properties as used for signal reconstruction, however, at the expense of some additional computational cost.

CS based solutions for detection problems in WSNs  are  summarized   in Table \ref{tab_detection}.

\subsection{Compressive Classification}
CS based classification work treated thus far in the literature  has mainly   dealt  with the temporal data. A classification problem with $C$ classes  can be formulated as a multiple hypothesis testing problem with $C$ hypotheses. Let the observation vector at the $j$-th node under the $i$-th hypothesis be
\begin{eqnarray}
\mathcal H_i: \mathbf x_j = \mathbf s_j^{(i)} + \mathbf v_j
\end{eqnarray}
for $j=1,\cdots,L$ and $i=1,\cdots,C$ where $\mathbf s_j^{(i)} \in \mathbb R^N$ is the signal of interest under $\mathcal H_i$. If all the nodes transmit their length $N$ observation vectors  to the fusion center, the fusion center makes the classification decision based on $\mathbf x = [\mathbf x_1^T, \cdots, \mathbf x_L^T]^T$ (or its noisy version).
With CS, the received compressed signal vector at the fusion center from the $j$-th node is given by
\begin{eqnarray}
\mathbf y_j = \mathbf A_j \mathbf x_j + \mathbf w_j\label{obs_classi}
\end{eqnarray}
where $\mathbf w_j$ is the noise vector at the fusion center. Classification is then performed using $\mathbf y = [\mathbf y_1^T, \cdots, \mathbf y_L^T]^T$ instead of $\mathbf x=[\mathbf x_1^T, \cdots, \mathbf x_L^T]^T$.

The specific classification method and the impact of the compression on the performance depend on how the signals $\mathbf s_j^{(i)}$'s and the noise vectors $\mathbf v_j$'s are modeled.
For example, when $\mathbf s_j^{(i)}$'s are deterministic and known, and elements of each $\mathbf v_j$ are iid Gaussian with mean  zero and variance $\sigma_v^2$, the maximum likelihood classifier decides the true hypothesis (class) to be
\begin{eqnarray}
\hat i &=& \underset{i}{\arg\max}~p(\mathbf x | \mathcal H_i)\nonumber\\
&=&\underset{i}{\arg\min} ~\sum_{j=1}^L ||\mathbf x_j - \mathbf s_j^{(i)}||_2^2.
\end{eqnarray}
based on $\mathbf x$.
With compressed data in \eqref{obs_classi} and assuming that each projection matrix is an orthoprojector, the ML classifier reduces to
\begin{eqnarray}
\hat i &=& \underset{i}{\arg\min} ~\sum_{j=1}^L ||\mathbf y_j - \mathbf A_j\mathbf s_j^{(i)}||_2^2.
\end{eqnarray}
Thus, as discussed in \cite{Davenport_SPIE2007} (for the case of $L=1$),  classification performance with compressed and uncompressed data depends only on how the distance measures are distorted by random projections. In particular, when $\mathbf A_j$'s are orthoprojectors, the distance between any two points in the compressed domain is reduced by approximately a factor of $M/N$ compared to that with compared to uncompressed data.  In \cite{Wimalajeewa_TSP12}, the authors have established the relationships with several probabilistic distance measures with compressed as well as uncompressed data considering Gaussian as well as non-Gaussian distributions for uncompressed data. The distance measures can be used to evaluate how good the compressive classification is.  In \cite{Vila-Forcen_CISS08}, design of  projection matrices to improve the  compressive classification performance  is discussed. Nonparametric approaches for classification exploiting CS have been discussed  in \cite{Wright_TPAMT2009,Sainath_ICASSP2010,Huang_NIPS2006} although not directly focusing on sensor data.

\subsection{Compressive Estimation}
Estimation is another important task performed  by sensor networks in addition to detection and classification. In this section, we describe recent work on CS based estimation. CS in fact deals with an estimation problem using an underdetermined linear system to recover  the projected high dimensional sparse  signal.  However, there are several applications where we are interested in estimating some parameters or functions without complete signal reconstruction. {We first focus on the use of CS for  parameter estimation in general and then discuss the CS based location estimation (or source localization)  problem.  }

\subsubsection{Compressive parameter estimation}
Instead of estimating the complete projected signal, it is of interest to explore how CS can be used to estimate a parameter (or a set of parameters) that govern the high dimensional signal. These types of problems arise in several applications such as time delay estimation \cite{Cao_SPL2017} and frequency estimation for a mixture
of sinusoids \cite{Fannjiang_SJIS2012,Duarte_ACHA2013,Ramasamy_TSP14}. Sparsity of the signal
is not a necessary requirement in such problems. Similar to the detection and classification problems discussed earlier, we need to  understand how much information is
retained in the compressed domain so that a reliable estimator can be obtained.

In  general form, consider  that the signal of interest $\mathbf x_j\in \mathbb R^N$ at the $j$-th node  is parameterized by some $K$-dimensional parameter vector $\boldsymbol\omega=[\omega_1, \cdots, \omega_K]^T$ with $K<N$. The compressed observation  vector at the $j$-th node \eqref{smv_model_noise} can be rewritten as
\begin{eqnarray}
\mathbf y_j = \mathbf A_j \mathbf x_j(\boldsymbol\omega) + \mathbf v_j.\label{Obs_est}
\end{eqnarray}
In order to quantify the information retained in the compressed domain, the authors in \cite{Ramasamy_TSP14} evaluated the impact of the random projections  on the estimation error. In particular, they have shown that the Cram$\grave{e}$r-Rao Lower Bound (CRLB), which  sets a
lower bound on the variance of any unbiased estimator, is increased approximately by a factor of $M/N$ with compressed data compared  to its uncompressed counterpart  when the noise is iid Gaussian. However, since the CRLB is a function of SNR, the performance of the  CS based estimation can still be very close to that with uncompressed data when SNR is large. Estimation of time-difference-of-arrival (TDOA) using compressed data via the ML estimation method has been discussed in \cite{Cao_SPL2017} without signal reconstruction. The TDOA estimates computed in the compressed domain can be used for source localization. These types of approaches are important when the sensors generate huge amounts of data and transmitting such large datasets is prohibitive due to communication constraints.

Another interesting formulation of compressive estimation is to estimate  a  function  of the data based on compressed measurements as discussed in \cite{davenport_JSTSP10} where  complete  signal reconstruction is not necessary. In this  framework with a SMV, the goal is to estimate a function of $\mathbf x$, $f(\mathbf x)$, based on $\mathbf y$ in (\ref{smv_model_noise}). In \cite{davenport_JSTSP10}, the authors have  considered  the case where $f(\mathbf x) = \langle \mathbf g, \mathbf x\rangle$ where $\mathbf g\in \mathbb R^N$ where $\langle \cdot \rangle$ denotes the inner product. The existing compressive parameter estimation techniques  mostly focus on the SMV case. However,  extensions to the MMV case are  worth investigating so that they are applicable for WSN applications.

\begin{table*}
\renewcommand{\arraystretch}{1.3}
 %if using array.sty, it might be a good idea to tweak the value of
% \extrarowheight as needed to properly center the text within the cells
\caption{CS based solutions for parameter  estimation and source localization   in WSNs}
\label{tab_localization}
\centering
% Some packages, such as MDW tools, offer better commands for making tables
% than the plain LaTeX2e tabular which is used here.
\begin{tabular}{|l|l|l|}
\hline
Problem & Features & References \\
\hline
\hline
Parameter estimation   & Framework for estimating deterministic (scalar/vector)  parameters    & \cite{Ramasamy_TSP14}\\
$~$ &  based on compressed data, derive CRLB, SMV & \\
$~$& Estimation of TDOA based on compressed data, use ML approach    & \cite{Cao_SPL2017}\\
\hline
Source localization  & SR with  SMV, MMV,  use  $l_1$ norm minimization  & \cite{Malioutov_TSP2005,Liu_IJDSN2012}\\
$~$ & SR with additional pre-/post- processing,  use  $l_1$ norm minimization & \cite{Feng_Globecom2009}\\
$~$ & SR  with MMV,   use OMP  & \cite{Nikitaki_ESPF2011}\\
   & SR  with  SMV,  use GMP  & \cite{Zhang_Infocom2011}\\
   $~$ & SR   MMV,  use GMP& \cite{Lagunas_ICASSP2016}\\
$~$ &SR  with  SMV,  use VBEM   & \cite{Sun_COM2017,Sun_COML2017}\\
$~$ & SR with  SMV,  use iterative BCS& \cite{Xiahou_IJDSN2015}\\
$~$ & Signal sparsity  and spatial SR with MMV,  distributed BCS & \cite{Cevher_IPSN09}\\
\hline

\end{tabular}
%\begin{enumerate}
%\item
%  \end{enumerate}
\end{table*}

\subsubsection{Localization}\label{sec_localization}
{Source localization has been an active research area for a long time which has applications in diverse fields. In the literature, the source localization problem has been  treated using  both parametric as well as nonparametric approaches. In a parametric framework, one of the commonly considered methods   is the ML  method which shows excellent statistical properties. However, the ML approach  is in general computationally  complex since obtaining a closed-form solution is difficult. Nevertheless, there have been some approaches proposed to exploit CS in estimating parameters such as the TDOA in the ML framework using compressed data  as discussed in the previous subsection which are used to perform source localization \cite{Cao_SPL2017}.}

Among several other techniques for source localization, sparse  representation (SR) based source localization has attracted  much attention over the years due to its capability of achieving super-resolution compared to other suboptimal localization methods \cite{Malioutov_TSP2005}.  The problem  of SR based source  localization has a similar form as considered for sparse event monitoring in Section \ref{sec_det_sparse_event}. In particular, with redefined notation, let there be $K$ sources and $M$ sensors. The emitted signals by the $K$ sources are given in   $\mathbf s=[s_1, \cdots,s_K]^T$. Further, let $\mathbf r=[r_1, \cdots,r_K]^T$ denote the source locations. The received signals at the $M$ sensors can be represented  as
 \begin{eqnarray}
 \mathbf y =  \boldsymbol\Psi(\mathbf  r)\mathbf s + \mathbf v\label{SR_0}
 \end{eqnarray}
 where the $(i,j)$-th element of the $M\times K$ array manifold matrix $\boldsymbol\Psi(\mathbf  r)$ contains the delay and gain information from the $j$-th source located at $r_j$  to the $i$-th sensor.
 In the SR framework, (\ref{SR_0}) is represented as an overcomplete representation by constructing a fine grid, to account for all possible source locations,  over the region of interest. Let $\tilde r = [\tilde r_1, \cdots, \tilde r_L]^T $ denote the grid locations where $L >> K$. Then, the matrix $\boldsymbol\Psi(\mathbf  r)$ is redefined as a $M\times L$ matrix whose $(i,j)$-the element corresponds to the gain and delay information between the source location $\tilde r_j$ and the $i$-th sensor. The signal field is given in $\tilde{\mathbf s}$ where the $n$-th element of $\tilde{\mathbf s}$ is nonzero if the $n$-th location has an active source while it is zero otherwise. With this representation, \eqref{SR_0} can be rewritten as
  \begin{eqnarray}
 \mathbf y =  \boldsymbol\Psi(\tilde{\mathbf  r})\tilde{\mathbf s} + \mathbf v\label{SR_1}
 \end{eqnarray}
 where $\tilde{\mathbf s}$ is a sparse vector. In this formulation, the sparse localization problem reduces to estimating a sparse signal  $\tilde{\mathbf s}$ based on $\mathbf y$ where the elements of the $ \boldsymbol\Psi(\tilde{\mathbf  r})$ are determined by the particular signal model under consideration.

%\subsection{Localization and tracking}
 The works in \cite{Malioutov_TSP2005,Liu_IJDSN2012} have used  $l_1$ norm minimization to impose sparsity to solve the source localization problem with SMV as well as MMVs.  {It has been shown that the SR based approach outperforms the existing localization techniques such as MUSIC in terms of the resolution, robustness to noise, number of time samples needed, and existence of dependence among sources.  However, the computational complexity of the $l_1$ norm based approach is quite significant compared to other localization approaches.  }  In \cite{Feng_Globecom2009}, RSSI-based source localization problem has been considered using a similar  spatial grid model used in  \eqref{SR_1}  employing  $l_1$ norm minimization.  Compared to the work in \cite{Malioutov_TSP2005,Liu_IJDSN2012},  pre-processing is used to induce incoherence needed
in  CS theory, and post-processing is included to compensate
for the spatial discritization  caused by the grid assumption in \cite{Feng_Globecom2009}.
Greedy techniques  for SR based source localization have been utilized in \cite{Zhang_Infocom2011,Nikitaki_ESPF2011,Lagunas_ICASSP2016} which have reduced computational complexities compared to $l_1$ norm based approaches.
 {In \cite{Zhang_Infocom2011}, a greedy   matching
pursuit algorithm (GMP),  has been proposed for  target counting and localization jointly using the SR  framework. It has been shown that the GMP algorithm is capable of providing a good trade-off between the localization performance  and the computational complexity compared to other SR-based and non SR-based approaches for localization. The GMP algorithm has been extended to take MMVs into account in \cite{Lagunas_ICASSP2016} with RSSI measurements. }
 In \cite{Nikitaki_ESPF2011}, OMP has been employed  exploiting the SR framework  using RSSI measurements. {While promising results have been illustrated compared to non-SR based localization approaches, performance comparison with other SR-based approaches  is missing in \cite{Nikitaki_ESPF2011}. }
 Bayesian algorithms for SR based source localization have been explored in \cite{Sun_COM2017,Sun_COML2017,Xiahou_IJDSN2015,Cevher_IPSN09}.   A multiple
target counting and localization framework using the variational
Bayesian expectation-maximization (VBEM)  algorithm has been developed in \cite{Sun_COM2017,Sun_COML2017}.  {It has been shown that, in contrast to $l_1$ norm minimization based and greedy algorithms, VBEM algorithm is more robust in localizing off-grid targets.} In \cite{Xiahou_IJDSN2015}, the authors have  proposed an adaptive BCS algorithm which  selects the number of sensors to be participated in the localization process based on the feedback of estimated variances of noise at different nodes.
  In  \cite{Cevher_IPSN09},  a distributed  source localization scheme have been proposed combining SR in spatial domain and signal sparsity at each node to reduce sampling cost. The algorithm has further been extended to take only 1-bit measurements of the compressed data in \cite{Cevher_IPSN09}.

{A summary of  CS based general parameter estimation and SR based solutions for  source localization is given in Table \ref{tab_localization}.   When analyzing  existing SR based source localization techniques, simulation results have been  provided in most of the work to analyze advantages and disadvantages over the performance and computational cost of different algorithms. However, theoretical analysis under  practical conditions such as the presence  of MMV, mismatches  of spatial grid, and stability of estimates is  lacking in most of the work which  is worth further investigating.}

\subsection{Sparsity Aware Sensor Management}
 Sensor management, also dubbed as \emph{sensor censoring}, is identified as one of the solutions for energy limitations in WSNs \cite{Joshi_TSP2009}. This involves in selecting  the best subset of sensors
that guarantees a certain inference  performance.  Finding the optimal solution for this problem is intractable in general and a wide class of sub optimal approaches, that fall into convex optimization, greedy and heuristic approaches,  have been proposed in the literature \cite{Zhao_TSP2002,Zuo_icassp2007,Carmi_Auto2013,Joshi_TSP2009}. The sensor selection problem has gained much attention lately  in the context of sparsity aware processing since efficient algorithms can be developed benefitting from the  inherent sparsity in many WSN applications.  In  this subsection, we review the sensor management work that exploits sparsity aware techniques. Sensor management for jointly sparse signal recovery in WSNs is addressed in \cite{Wu_TCOM18} when sparse measurement vectors are used to compress  observations at distributed nodes. The goal is to design a ternary protocol at the sensors to decide whether  to transmit the compressed measurement vector, transmit a 1-bit  hard decision, or not transmit  based on a error criterion defined in terms of the  overlap between the signal support and the support of the measurement vector.  Centralized and distributed algorithms for sparsity aware sensor selection for the problem of estimation with a linear model have been  proposed in \cite{Jamali-Rad_SPL2014}. In \cite{Liu_TSP2015}, the sensor selection problem  for linear estimation has been  formulated as a sparsity aware optimization problem considering two types of sensor collaboration; information constrained  and  energy constrained collaborations. The authors have employed  reweighted $l_1$ norm based ADM and the bisection algorithm to solve the optimization problem.  In \cite{Chepuri_TSP2015}, the sensor selection problem has been formulated as the design of a sparse vector  considering a  nonlinear estimation framework. In \cite{Masazade_SPL2012},  sensor selection for
target tracking  has been
considered  in which the selection is performed by designing
a sparse gain matrix.  A probabilistic sensor management scheme for target tracking has been  proposed in \cite{nianxia_tsp14} where the MAC model with distributed sparse random projections  as discussed in \ref{sec_spatial} is   used to compress the spatially sparse data. In \cite{Chen_TVT2016}, the authors have exploited the CS framework to develop a node selection
mechanism  for compressive sleeping WSNs to improve the performance
of data reconstruction accuracy, network lifetime, and
spectrum resource usage.

{In Sections \ref{sec_data_gathering} and \ref{sec_inference}, exploitation of  CS  with and/or without  reconstruction has been discussed categorizing  the work by specific  task of interest in WSNs. The following section is devoted to discuss approaches  to cope with practical aspects when applying CS in WSNs. It is noted that, different aspects discussed below are  applied to any given task in general. }

\section{Practical Aspects}\label{sec_practicalCon}
As discussed in the Introduction section, for  CS based solutions  to be practical in different  WSN applications, it is important  to investigate how well the CS based techniques perform  in  practical communication networks; i.e., in the presence of practical issues    such as channel impairments, security related issues and measurement quantization.

\subsection{Impact of Fading  Channel Impairments}\label{sec_fading}
In practical communication networks, the communication
channels between the sensor nodes and the fusion center undergo
fading. The presence of fading affects the recovery/inference  capabilities of CS based techniques. For example, consider the CWS framework discussed  in Section \ref{sec_spatial}. In the presence of fading, the observation at the fusion center after the $i$-th MAC transmission, \eqref{MAC_AWGN_i}, can be rewritten as
\begin{eqnarray}
y_i = \sum_{k=1}^L  h_{i,k}\mathbf A_{i,k} x_k + v_i\label{MAC_Fading}
\end{eqnarray}
for $i=1,\cdots, M$ where $h_{i,k} $ is the fading coefficient for
the channel between the $k$-th sensor and the fusion center during
the $i$-th transmission.  In vector notation, \eqref{MAC_Fading} can be expressed as
\begin{eqnarray}
\mathbf y = \mathbf B \mathbf x + \mathbf v\label{MAC_Fading_v}
\end{eqnarray}
where $\mathbf B= \mathbf H \odot \mathbf A$ where $\mathbf H$ is the $M\times L$ channel matrix with  $\mathbf H_{i,k} = h_{i,k}$ and $\odot$ denotes the Hadamard (element-wise) product. With the model  in \eqref{MAC_Fading_v}, the effective measurement matrix, $\mathbf B$,  has different properties compared to that of $\mathbf A$. Due to this, the recovery guarantees developed assuming AWGN channels may not be valid in the presence of fading
since it leads to inhomogeneity and non Gaussian statistics
in measurement matrices. In \cite{Yang_TSP13,Yang_icc13}, the problem of sparse signal
recovery based on \eqref{MAC_Fading_v} has been  addressed where
the authors provide \emph{uniform} recovery guarantees \cite{Rauhut_J2010} based on
RIP when $\mathbf A$ is chosen as  a sparse Bernoulli
matrix \eqref{sparse_Bern}.  The authors in \cite{Wimalajeewa_TSIPN15} have extended the work in \cite{Yang_TSP13} to derive \emph{nonuniform } recovery guarantees \cite{Rauhut_J2010} in the presence of Rayleigh fading in which the sparse Gaussian matrix \eqref{sparse_Gauss} is used for distributed compression. In particular, the heterogeneousness and the heavy tailed behavior of the projection matrix due to fading  requires  the sensors  to take slightly more measurements   than that required with AWGN channels to guarantee reliable recovery \cite{Wimalajeewa_TSIPN15}.
The robustness of the CS framework in the presence of nonideal channels has  also been  discussed in \cite{Chen_TCAS2013,Chae_MICOM2010}. The impact of fading and noisy communication channels in random access CS, an
efficient method for data gathering \cite{Fazel_JSAC2011}, has been investigated in \cite{Fazel_TWC13}.

Another practical aspect to be dealt with is to understand the impact of security issues in WSNs on the  CS framework.

\subsection{Physical Layer Secrecy Constraints}
 In WSNs, transmissions by distributed  nodes may be
observed by eavesdroppers. Further, the network may operate under malicious and Byzantine attacks. Thus, the secrecy of a networking  system
against eavesdropping attacks is of utmost importance \cite{Choi_PIMRC2013}. In a
fundamental sense, an  eavesdropper can be  selfish and malicious, to compromise the
secrecy of a given inference network.
In the recent past, there has been a significant  interest in
the research community in addressing eavesdropping attacks on
distributed inference networks.  However, while  there  are a few recent works \cite{Agrawal_PITW2011,Rachlin_Allerton2008,Barcelo_TIFS2014,Dautov_ICNC2013,
Kailkhura_TSP16,Kailkhura_WCL16}, a detailed  study  with respect to CS based techniques
has not yet been done thus far.

In \cite{Agrawal_PITW2011,Rachlin_Allerton2008}, the performance limits
of secrecy of CS  have been  analyzed. The amplify-and-forward CS scheme is introduced in \cite{Barcelo_TIFS2014} as a physical layer secrecy solution for WSNs. The authors  have studied the  secrecy performance  against a passive eavesdropper agent composed of several malicious and coordinated
nodes.  A  CS encryption framework for
resource-constrained WSNs has been  proposed in \cite{Dautov_ICNC2013}. The authors  establish a secure sensing matrix, which is  used
as a key, by utilizing the intrinsic randomness of the wireless
channel. In \cite{Kailkhura_TSP16}, the authors have considered the distributed  compressive detection problem in the presence of eavesdroppers. The authors have derived   the optimal system parameters to maximize detection performance at the fusion center
while ensuring perfect secrecy at the eavesdropper. In \cite{Kailkhura_WCL16}, the problem of  designing  measurement matrices for compressive detection  so that the detection performance of
the network is maximized while guaranteeing a certain level of
secrecy has been  discussed.

{Next, we discuss  the impact of measurement quantization on  CS based processing. }

\subsection{Sparse Signal Reconstruction  with Quantized CS}\label{sec_quantization}
Compression  achieved via random projections at the sampling  stage is desirable in resource constrained WSNs. However, further compression/quantization  of the compressed measurements may be   necessary   in many WSN applications operating  under severe  resource constraints. Coarse quantization reduces the bandwidth requirements  and computational costs at local nodes, and thus,  is well motivated for practical WSNs.  The traditional CS framework with real valued measurements has been   extended to take quantization effect into account. In particular, one of the driving forces behind the development of a quantized CS framework is the motivation to further reduce the amount of information to be communicated in resource constrained WSNs.  Consider the compressed observation model as given in \eqref{smv_model_nonoise} or \eqref{smv_model_noise}. With quantized CS, one has access to $\mathbf z = \mathcal Q(\mathbf y)$ instead of $\mathbf y$ where $\mathcal Q(\cdot)$ is an entry-wise (scalar)  quantizer which maps real valued measurements to  a discrete quantized alphabet of size $Q$. In particular,  each  element of $\mathbf y$ is quantized into $Q$ levels so that
\begin{eqnarray}
z_i=\left\{
\begin{array}{ccc}
0, ~&~ if~ \tau_0 \leq y_i < \tau_1\\
1, ~&~ if~ \tau_1 \leq y_i < \tau_2\\
.\\
.\\
Q-1, ~&~ if~ \tau_{Q-1} \leq y_i < \tau_Q
\end{array}\right.\label{quant1}
\end{eqnarray}
for $i=1,2,\cdots,M$, where $\tau_0, \tau_1,\cdots, \tau_Q$ represent the quantizer thresholds with
$\tau_0 = -\infty$ and $\tau_Q = \infty$. With this approach, $\lceil{\log_2 Q\rceil}$ bits per measurement are required to transmit each element of $\mathbf y$. In the special case with a 1-bit quantizer,   the measurements are quantized into only two levels such that $Q=2$.  One example under this special case is to use only the
sign information of the compressive measurements \cite{Boufounos_CISS2008,Gupta_ISIT2010,Boufounos_ICASSP2010,
Jacques_TIT13,Boufounos_Asilomar09}.
 More specifically, the  1-bit CS scheme first proposed in \cite{Boufounos_CISS2008} with sign measurements is given by
 \begin{eqnarray}
z_i=\left\{
\begin{array}{ccc}
1, ~&~ \mathrm{if}~   y_i \geq 0\\
-1, ~&~ \mathrm{otherwise}\\
\end{array}\right.\label{quant_1bit}
\end{eqnarray}
for $i=1,2,\cdots,M$. Equivalently, we may express \eqref{quant_1bit} by
\begin{eqnarray}
\mathbf z = \mathrm{sign}(\mathbf y)\label{1-bit_sign}
\end{eqnarray}
where $\mathbf z=[z_1, \cdots, z_M]^T$, and $\mathrm{sign}(\cdot)$ denotes the element-wise sign operation. Sparse signal processing
 with 1-bit quantized CS is attractive since 1-bit CS techniques are robust under different kinds of non-linearties applied to measurements and have less sampling complexities at the hardware level because the quantizer takes the form of  a comparator \cite{Boufounos_CISS2008,Boufounos_ICASSP2010}.

Now, the goal is to perform sparse signal recovery or solve inference tasks based on $\mathbf z$ instead of $\mathbf y$. There are several factors that need to be taken into account  when evaluating the performance and developing algorithms  with quantized CS especially as applicable to resource constrained WSNs;
\begin{itemize}
\item Quantization introduces nonlinearity. Thus, the algorithms  and the recovery guarantees developed for sparse signal processing with  real valued CS may not be directly applicable for quantized CS. This provides the motivation  to develop quantization schemes and reconstruction algorithms so that the performance with quantized CS is very close to  (or even  better than) that with the real valued CS.
    \item Coarse quantization can  reduce the ability of signal reconstruction or performing inference tasks compared to real valued CS. This  provides the impetus   to take more measurements to compensate for the loss due to quantization. Thus, investigation of    the trade-off between the cost for quantization and the cost for sampling is important.
        \item  In distributed networks, MMVs appear naturally. This motivates us to extend the quantized CS based processing set-up to the distributed/decentralized  settings with different communication architectures.
\end{itemize}
Over the past few years, there have been several research efforts that aim to consider quantized CS in different contexts.  In the following, we first review the work on 1-bit CS.
\subsubsection{Algorithm development and performance guarantees with 1-bit CS}
Most of the early work related to 1-bit CS considers the SMV case.
In \cite{Boufounos_CISS2008}, the  authors have introduced the 1-bit CS  problem with sign measurements  (\ref{1-bit_sign}) for  the noiseless case (i.e.,  $\mathbf y$ is as in \eqref{smv_model_nonoise}). The authors  have developed  an optimization based algorithm for sparse signal recovery using a variation of the fixed point continuation
(FPC) method \cite{Hale_TR2007}. In particular, they have considered  solving
\begin{eqnarray}
 \underset{\mathbf x} {\min} ||\mathbf x||_1 ~ \mathrm{such} ~\mathrm{that} ~ \mathbf Z \mathbf A \mathbf x \geq 0 ~ \mathrm{and} ~ ||\mathbf x||_2=1\label{opt_sign}
  \end{eqnarray}
where $\mathbf Z=\mathrm{diag}(\mathbf z)$. It is noted that, for an $N\times 1$ vector $\mathbf x$,  $\mathrm{diag}(\mathbf x)$ denotes a $N\times N$ diagonal  matrix in which the main diagonal is composed  of elements of  $\mathbf x$.
The authors have shown that the recovery performance can be significantly improved with the proposed  algorithm compared to employing classical CS algorithms when the measurements are quantized to 1-bit. One problem in \eqref{opt_sign} is that it requires the solution of  a non-convex optimization problem. In \cite{Laska_TSP2011}, the authors have presented a provable optimization algorithm to solve \eqref{opt_sign}. A convex formulation of the signal recovery problem  from  1-bit CS  has been presented in \cite{Plan_CPAM2013} which solves for
 \begin{eqnarray}
 \underset{\tilde{\mathbf x}} {\min} ||\tilde{\mathbf x}||_1 ~ \mathrm{such} ~\mathrm{that} ~ \mathrm{sign}( \mathbf A \tilde{\mathbf x} ) = \mathbf z ~ \mathrm{and} ~ ||\mathbf A\tilde{\mathbf x}||_1=M. \label{opt_sign_convex}
  \end{eqnarray}
  It has been shown in \cite{Plan_CPAM2013} that, (\ref{opt_sign_convex}) can provide an arbitrarily accurate
estimation of every $k$-sparse signal $\mathbf x$   with high probability  when $M = \mathcal O(k \log^2(N/k))$ 1-bit measurements. According to their results, the required number of CS measurements  matches the known results with real valued CS up to the  exponent 2 of the logarithm and up to an absolute constant factor.   In \cite{Fang_SP2014,Shen_ChinaSIP2013},  the log-sum penalty function,  which has the potential
to be much more sparsity-encouraging than the $l_1$ norm,   has been  used for sparse signal
recovery with 1-bit CS. They have developed an iterative reweighted algorithm which consists of solving a
sequence of convex differentiable minimization problems for sparse signal recovery.

Greedy and iterative algorithms developed for classical CS have been extended to the 1-bit CS case in   \cite{Boufounos_Asilomar09,Jacques_TIT13,North_arXiv15}. Modification of the CoSAMP algorithm for the 1-bit case  to produce Matching
Sign Pursuit (MSP) has been  presented in \cite{Boufounos_Asilomar09}. In \cite{Jacques_TIT13}, an extension of the  IHT algorithm  with 1-bit CS, called binary IHT (BIHT) has been  developed. The BIHT algorithm to incorporate additional information on the partial support has been  considered in \cite{North_arXiv15}.   The authors in \cite{Xu_JSMTE2014} have  presented a Bayesian approach for
signal reconstruction with  1-bit CS, and analyzed its typical performance
using statistical mechanics.  In \cite{Kamilov_SPL2012}, a Bayesian algorithm based on generalized approximate message passing (GAMP) has been  developed.

Recovery guarantees and consistency
of the estimator for both Gaussian and sub-Gaussian random measurements have been established in \cite{Chen_AISTATS2015} for 1-bit CS using the recently proposed  k-support norm \cite{McDonald_NIPS2014}.   In \cite{Zhang_ICML2014}, the sample complexity of vector recovery using 1-bit CS has been  discussed.
In a recent work presented in \cite{Baraniuk_TIT2017}, the authors have shown  that 1-bit measurements allow for exponentially decreasing error with adaptive thresholds. This framework is slightly different from the 1-bit CS model discussed with sign measurements in \cite{Boufounos_CISS2008}.  More specifically, the $i$-th element of $\mathbf z$ is given by
 \begin{eqnarray}
z_i=\mathrm{sign}(y_i - \nu_i) = \left\{
\begin{array}{ccc}
1, ~&~ \mathrm{if}~   y_i \geq \nu_i\\
-1, ~&~ \mathrm{if} ~ y_i < \nu_i\\
\end{array}\right.\label{quant_1bit_adaptive}
\end{eqnarray}
for $i=1,\cdots,M$. This scheme  allows the
quantizer to be adaptive, so that $\nu_i$ in \eqref{quant_1bit_adaptive} of the $i$-th entry may depend on the $1\mathrm{st}, 2\mathrm{nd},\cdots, (i-1)\mathrm{st}$ quantized measurements. Adaptive one-bit quantization has also been considered in \cite{Fang_SP2016} where the authors have shown that  when the number of one-bit measurements is
sufficiently large,  the sparse signal
can be recovered with a high probability with a bounded error. The error bound is
linearly proportional to the $l_2$ norm of the difference
between the thresholds and the original unquantized
measurements.   In another  recent work by Knudson et.al. in \cite{Knudson_TIT2016}, it has been shown that  norm recovery is possible with sign measurements of the form $\mathrm{sign}(\mathbf A\mathbf x + \mathbf b)$ for random $\mathbf A$ and fixed $\mathbf b$ which is impossible with $\mathrm{sign}(\mathbf A\mathbf x )$.

\subsubsection{Algorithm development and performance guarantees with quantized CS}
The authors in \cite{Zymnis_SPL2010,Wimalajeewa_JSTSP12,Jacques_TIT2011,dai_ISIT2009} have considered  the sparse signal/support recovery problem with a given scalar qunatizer where 1-bit CS appears as a special case. In scalar quantization, each element of $\mathbf y$ is quantized independently as shown in \eqref{quant1}. In \cite{Zymnis_SPL2010,Wimalajeewa_JSTSP12}, several algorithms and performance bounds have been  derived considering a SMV.  More specifically,  Zymnis et.al. in \cite{Zymnis_SPL2010} have developed two algorithms for sparse signal recovery with quantized  CS by minimizing a  differentiable convex function plus
an $l_1$ regularization term.   In \cite{Wimalajeewa_JSTSP12}, the  authors have investigated the minimum number of CS measurements required for support recovery with a given  quantizer  (including 1-bit CS)  with a SMV specifically focusing on support recovery. The effect of quantization has  further been analyzed in \cite{Ardestanizadeh_ISIT2009,Boufounos_SampTA2011,Jacques_TIT2011,Boufounos_TIT2012,dai_ISIT2009}. The effects of precision
in the measurements have been analyzed in  \cite{Ardestanizadeh_ISIT2009} by considering  the syndrome decoding
algorithm for Reed-Solomon codes when applied in the
context of compressed sensing as a reconstruction algorithm
for Vandermonde measurement matrices. Universal  scalar quantization with
exponential decay of the quantization error as a function of the
oversampling rate has been  considered in \cite{Boufounos_TIT2012}. In particular, the author has shown that non-monotonic quantizers achieve exponential error decay in the oversampling
rate using consistent reconstruction.  However, reconstruction
from such a quantization method is not straightforward, and the same author has proposed a practical algorithm  for reconstruction using a hierarchical quantization approach in \cite{Boufounos_SampTA2011}.  In \cite{Jacques_TIT2011}, the authors have presented  a variant of basis pursuit denoising, based on  $l_p$
norm rather than using the $l_2$ norm. They have  proved that the performance of the proposed  algorithm  improves with larger $p$. In \cite{dai_ISIT2009}, an adaptation
of basis pursuit denoising and subspace sampling has been  proposed
for dealing with quantized measurements.

Design and analysis of vector quantizers (VQ) with CS measurements have been considered in \cite{Shirazinia_ISITA2012,Shirazinia_TSP2014}.  In \cite{Shirazinia_ISITA2012}, the authors have  addressed the design of VQ  for block sparse signals
using block sparse recovery algorithms. Inspired by the Gaussian mixture model (GMM)
for block sparse sources, optimal rate allocation has been
designed for a GMM-VQ which aims to minimize quantization
distortion. In \cite{Shirazinia_TSP2014},  optimum joint source-channel VQ schemes have been  developed for CS measurements. Necessary conditions for the optimality of vector quantizer  encoder-decoder
pair with respect to end-to-end MSE have been  derived.

\subsubsection{Distributed and decentralized solutions with quantized CS}
While most of the existing works on quantized CS are  restricted to the  SMV case, there are a few recent works that have extended  the quantized CS framework to the multiple sensor setting. In \cite{Chen_WCL15}, sparse signal reconstruction using 1-bit CS has been considered modeling communication between sensors and the fusion center via a binary symmetric channel (BSC). Instead of using the sign measurements as in \eqref{quant_1bit}, local binary quantizers are designed to cope with the bit flipping caused by  BSC. Several existing CS algorithms developed for real valued CS  have been been extended to the quantized CS  setting  in \cite{Gupta_Asilomar2015,Kafle_TSIPN2017,Kafle_GlobalSIP2016} considering centralized as well as distributed/decentralized implementations.
 Solving the support recovery problem in a decentralized setting with 1-bit CS (sign measurements)  has been  considered  in \cite{Kafle_TSIPN2017,Kafle_GlobalSIP2016} where the authors have developed several  decentralized versions  of the BIHT  algorithm. In \cite{Leinonen_SPAWC2016}, the authors employed a distributed variable-rate quantized CS methodology for acquiring
correlated sparse sources in WSNs. Optimality
conditions that minimize a weighted
sum of the average MSE distortion for  signal recovery with complexity-constrained encoders have been derived.

\subsection{Other Issues}
{In addition to the practical issues discussed above, robustness of the CS scheme against   missing or erroneous information, node failures, and other types of errors has been investigated by several researchers. It has been shown  that the multi-hop data gathering scheme  proposed in \cite{haupt_SPM2008} employing randomized gossip algorithms is  robust to node failures and changes in network configurations.  In \cite{Chen_mobicom2014}, compression of spatio-temporal data with missing information and in the presence of anomalies has been addressed. As discussed in Section \ref{sec_matrix_completion}, the data  matrix has a low rank structure in the presence of  spatio-temporal sparsity.  When there are missing information and/or anomalies,  assumptions made in general CS theory may be violated  leading to inaccurate/degraded performance if they are not properly taken care of. By proper decomposition of the data matrix to take the  low-rank property, missing value interpolation and noise terms into account, the authors in \cite{Chen_mobicom2014} have developed a nuclear norm minimization based approach which is shown to be robust against missing information and anomalies. CS recovery techniques have  also been used in \cite{Kong_TPDS14} as a tool for estimating spatio-temporal  data in the presence of  missing information in WSNs  without employing any data compression as discussed in Section \ref{sec_matrix_completion}.  }

{While application of CS in WSNs   has a quite  rich literature as of now  along most of the categories discussed above,  there are still avenues for future research which  will be discussed next.  }
%\section{Testbed Implementation}

\section{Lessons Learned and Future Research Directions}\label{sec_future}
%While CS has a relatively rich literature as of  now, its implementation under practical communication  considerations as desired by WSNs has been discussed only during past few years.
With the emergence of new technologies such as IoT where   heterogeneous networking architectures need to be integrated to perform a variety of tasks, handling and integration of a large amount of  data  generated by the interaction of multiple factors/sensors  keeps continuously challenging. In particular, future WSNs are expected to integrate with a variety of other networks such as wireless mesh networks, Wi-Fi, and vehicular networks to make smart platforms for IoT applications. Understanding the role of CS, as a tool to  cope with the problem of data deluge,  is of great importance in making such applications realizable.

As reviewed in this paper specifically focusing on WSNs, there are several ways that CS techniques can be beneficial. While there has been   quite significant amount of work done during last several years to make CS based techniques practically implementable  in WSNs   under network resource constraints focusing on different problems,  there are  still challenges and open issues worth further investigation in the areas discussed in Sections \ref{sec_data_gathering}, \ref{sec_inference} and \ref{sec_practicalCon}.  In the following, we discuss such challenges and future research directions.

%\textcolor{blue}{\bf In particular, future WSNs are expected to integrate with a variety of other networks such as wireless mesh networks, Wi-Fi, and vehicular networks to make smart platforms for IoT applications.  Thus, there is a growing challenge in handling huge volumes of data generated  at unprecedented levels in diverse environments. Understanding the role of CS techniques in such environments needs further research.} In the following, we discuss such challenges and future research directions.
%Next, we discuss some future  directions for  research in CS based signal processing in resource constrained WSNs.
%\subsection{Improvements on the existing methodologies}
% Thus, there is still a growing interest in further investigating several issues in detail. In the following, we summarize  some of the potential areas worth investigating.

\subsection{Scalable Network Processing Exploiting Sparsity in Multiple Dimensions with  Heterogeneous Data}
As discussed in Section \ref{sec_data_gathering}, in CS based data gathering, the attention is mainly given to the case when  sparsity is defined with respect to a single vector; temporal  as in \ref{sec_temporal} and spatial  as in \ref{sec_spatial}.  To exploit spatio-temporal correlation in  data gathering,  there are  a few approaches as discussed in Section \ref{sec_spatio_temporal} which define  sparsity in 2-dimensions (2-D). Mainly,  the ideas of Kronecker CS and matrix completion \cite{Duarte_TIP2012,Candes_FCM2009} have been exploited under  restricted assumptions such as centralized processing. Further, when exploiting   Kronecker CS ideas,   reconstruction is performed after transforming  2-D (spatio-temporal) data to a single vector which requires  large memory and computational costs. Such requirements are  not desirable especially with on-line WSN applications. Thus, exploring approaches to reduce  computational requirements when exploiting spatio-temporal sparsity is useful in many potential applications. One direction of research interest is to consider  decentralized/cluster based  settings where only partial information is processed at multiple sinks or clusters so that the overall computational burden is distributed. Further, ideas developed in   \cite{Fang_SCIS2012,Wimalajeewa_Arx13} can be utilized to solve these  kind of problems in the matrix form with less computational cost  without using the vector form.

Beyond WSNs, when integrating different network architectures for future IoT applications, large amount of  data can be generated by the interaction of multiple factors/sensors  and thus can be intrinsically represented by higher order
tensors.   Thus, exploitation of low dimensional properties of high order tensors in an efficient manner is needed to better employ CS based compression techniques to get a variety of tasks done. In   CS theory, there has been  quite significant attention to generalize the CS framework for high dimensional multi-linear models \cite{Cichocki_SPM2015,Goldfarb_arxiv12}. Due to  the same reasons discussed as in Section \ref{sec_motivation_WSN}, the direct use of such techniques may not be desirable and extensions to cope with the  available network resources need to be explored.

\subsection{Distributed/Decentralized Processing  with Quantized CS}
While quantized CS (especially 1-bit CS), as discussed in Section \ref{sec_quantization} is appealing  for WSN applications to further reduce the communication cost, most of  the existing work on quantized CS is restricted to the SMV case.  Recently, the  works reported in \cite{Gupta_Asilomar2015,Kafle_TSIPN2017,Kafle_GlobalSIP2016} have extended the 1-bit CS framework to the multiple-sensor setting. However, its development is still in its infancy. Understanding the relationships among the network parameters, compression ratio (achieved via sampling with  CS) and the quantization levels (achieved via quantizing CS samples) focusing on decentralized processing in both data gathering (signal reconstruction) and inference perspectives is necessary to better utilize the quantized CS framework in potential WSN applications. Further, quantifying the performance difference in terms of reconstruction and inference when using coarse quantized CS and real valued CS would help understand the trade-off between  quantization and the performance.  Thus, further efforts are required to evaluate the benefits of quantized CS in WSNs.

\subsection{CS Based Inference with Multi-Modal Data}
When performing inference tasks from compressed  data without complete signal reconstruction, it is of importance  to understand how much information is retained in the compressed domain so that a reliable inference decision can be made. This depends on how the signal and noise are  modeled. When the (uncompressed) multisensor data is independent  and corrupted by additive Gaussian noise, performance metrics for CS based detection, classification, and parameter estimation have been  discussed in several recent works as reviewed in  Section \ref{sec_inference}.   Dependence is one of the common characteristics exhibited in  multiple sensor data which is hard to model  analytically with multimodal data unless the data is Gaussian.  Further studying CS based inference in terms of deriving performance metrics, developing communication efficient algorithms,  and designing   projection matrices in the presence of inter-modal and intra-modal dependence will enable the efficient use of CS in  many potential applications.   While there exist several recent works that exploit spatial (or inter-modal) dependence in   detection problems in the Bayesian CS framework  under restricted assumptions \cite{Wimalajeewa_TSP17,Wimalajeewa_CAMSAP2017}, CS based multi-sensor dependent data fusion especially in the presence of non-Gaussian pdfs  and  spatio-temporal dependence  is not well understood. Thus, exploitation of higher order dependence and structured properties of high dimensional data in  CS  based multi-sensor fusion  is worth investigating.

 In addition  to domain specific applications, when WSNs are used as a smart architecture  for IoT applications, it is expected  to process multi-modal  data efficiently in dynamic, and heterogeneous environments to perform  a variety of different tasks. Understanding how the CS framework can be beneficial  in an adaptive manner to perform multiple tasks under such environments is essential in the  realization   of future IoT  techniques.

\subsection{Further Developments Under Practical Considerations}
Understanding the robustness of the CS framework to practical systems in the presence of practical issues  such as fading channels, interference, non-Gaussian noise impairments, synchronization errors, quantization errors, link failures, and missing data     is important to make  CS based techniques useful  in WSNs.  As discussed in Section \ref{sec_practicalCon}, many desirable conditions  required to establish performance guarantees in the  standard CS framework may not be satisfied  in  the presence of such practical issues. While some recent work exists in this area, further investigation of approaches to rectify the adverse impact of such practical aspects on the overall  performance is needed.

\subsection{Testbed Experiments and Performance Evaluation}
% There has been    significant advances in modifying  and extending  theory and algorithms for CS based processing  under communication and energy constrained in WSNs.
{So far, most research on CS in WSNs is restricted to theoretical and algorithmic development.  There are  some  testbed experiments and demonstrations done to incorporate CS based techniques specifically focusing on data gathering. As discussed in Section \ref{sec_adaptive}, several testbed experiments analysis with real data  have been reported for CS assisted data gathering exploiting spatial sparsity for different applications  \cite{Chou_LCN2009,Wang_INFOCOM12,Aderohunmu_Sensor2015,Yin_Sensor2016}. Further, in \cite{Cheng_TWC13}, testbed experiments have been performed for CS based spatio-temporal data collection and recovery. However, in order to bridge the theory and practice,  further development is needed to validate the proposed techniques with real applications. In particular, testbed experiments  with  CS based techniques for a variety of applications with and/or  without complete signal reconstruction,   performance evaluation, and robustness analysis against   practical considerations with real data will be useful.}

\section{Conclusion}\label{sec_conclusion}
In this survey paper, our goal was to describe the research trends and recent work on the use of CS in a variety of  WSN applications. By discussing the motivational factors, we have identified several challenges that need to be addressed to enable  practical implementation of  CS based techniques in WSNs. To that end, we have reviewed the recent works that focus on developing centralized, distributed and decentralized solutions for data gathering and reconstruction exploiting temporal, spatial  as well as  spatio-temporal sparsity under communication resource constraints.  We then described the work addressing the   benefits of using the CS based techniques in solving several  inference problems including detection, classification, estimation and tracking as desired by many WSN applications. We have further provided a discussion on incorporating practical  considerations, such as channel fading, physical layer secrecy constraints  and quantization,  into the CS framework. Finally,  potential future research directions  in CS for resource constrained WSNs  have  been discussed. With this review paper, the readers are expected to gain  useful insights on the applicability of various CS techniques in solving a variety of WSN related problems involving high dimensional data.

\bibliographystyle{IEEEtran}
\bibliography{IEEEabrv,ref_SurveyP16}

\end{document}